\documentclass[english,12pt]{article}
\usepackage[T1]{fontenc}
\usepackage{enumitem}
\usepackage[utf8]{inputenc}
\usepackage{geometry}
\usepackage{adjustbox}
\usepackage{epigraph}
\usepackage{threeparttable}
\usepackage{amsmath}
\usepackage[plain]{fancyref}
\usepackage{microtype}
\usepackage{amsthm}
\usepackage{multirow}
\usepackage{array}
\usepackage{tabularx}
\usepackage{amssymb}
\usepackage{afterpage}
\usepackage{placeins}
\usepackage{bbm}
\usepackage{float}
\usepackage{setspace}
\usepackage{booktabs}
\usepackage[font={small}]{caption}
\usepackage{subcaption}
\usepackage{babel}
\usepackage{xcolor}
\usepackage{xurl}
\usepackage[colorlinks=false,
            allbordercolors={white}]{hyperref}

\usepackage[backend=biber,
            style=nature,
            autocite=superscript,
            isbn=false,
            eprint=false,
            url=false,
            clearlang=true,
            doi=false,
            date=year,
            hyperref=true,
            natbib=true]{biblatex} 

\AtEveryBibitem{%
  \clearfield{note}%
}      
  
\usepackage{mathpazo}
\usepackage{tcolorbox}

\usepackage{authblk} 
\usepackage{parskip}

\title{Simple contagion drives population-scale platform migration}

\author[1]{Dorian Quelle}
\author[2,3]{Frederic Denker}
\author[4]{Prashant Garg}
\author[1,5,$\ast$]{Alexandre Bovet}
\affil[1]{Department of Mathematical Modeling and Machine Learning, University of Zurich, Switzerland}
\affil[2]{Programme on Innovation and Diffusion, London School of Economics and Political Science, London, United Kingdom}
\affil[3]{Paris School of Economics, France}
\affil[4]{Department of Economics and Public Policy, Imperial College London, London, United Kingdom}
\affil[5]{Digital Society Initiative, University of Zurich, Switzerland}
\affil[$\ast$]{alexandre.bovet@uzh.ch}

\usepackage{graphicx}

\DeclareCaptionLabelFormat{sfiglab}{\textbf{Figure S#2}}
\DeclareCaptionLabelFormat{stablab}{\textbf{Table S#2}}

\begin{document}

\maketitle

{\bfseries Social media platforms mediate professional communication \autocite{gargPoliticalExpressionAcademics2025}, political expression \autocite{bailExposureOpposingViews2018,cinelliEchoChamberEffect2021}, and community formation\autocite{falkenbergPatternsPartisanToxicity2024a}, making the rare instances when users collectively abandon an incumbent platform particularly consequential. Strong network effects raise switching costs and strengthen incumbents' positions, making coordinated exit difficult \autocite{katzSystemsCompetitionNetwork1994,rietveldPlatformCompetitionSystematic2021}.
Homophily and shared shocks create correlated timing and clustering of adoptions that confound estimates of peer influence and obscure whether contagion is simple or complex \autocite{granovetterThresholdModelsCollective1978,aral2009distinguishing,shaliziHomophilyContagionAre2011,watts2002simple,centola2007complex}. Here we link 276{,}431 scholars on Twitter/X to their respective new profiles among the universe of all 16.7 million Bluesky accounts, tracked from January~2023 to December~2024, using a scalable, high-precision cross-platform matching pipeline. Exploiting exogenous variation from Brazil’s court-ordered suspension of Twitter/X and a dynamic matching design, we show that adoption is peer-driven, treatment effects are short-lived and dose-dependent, and contagion is simple, not complex. Three patterns characterize adoption and retention. Adoption concentrates among users deeply embedded in Twitter’s social graph. Public political expression predicts migration, consistent with homophilous inflows into a largely left-of-center Bluesky information space \autocite{quelleBlueskyNetworkTopology2025}. Early reconnection with prior contacts predicts longer tenure and engagement. While observational studies document clustered adoption consistent with peer influence \autocite{jeong2024exploring, cava2023drivers, heFlockingMastodonTracking2023, bittermann2025social}, causal identification at the population scale has remained elusive. Our findings provide the first population-scale causal evidence of peer influence in a social media platform migration by exploiting exogenous exposure variation in a natural experiment and using daily dynamic matching. Rather than the complex contagion mechanism often emphasized in the literature, contagion is predominantly simple \autocite{centola2007complex,guilbeault2021topological}. High-investment users are more likely to move, revising the view that sunk costs and incumbency anchor heavy users. Together, these patterns suggest that academics hedge platform risk rather than remain locked in \autocite{fang2021platform}. Our findings recast migration as a multi-homing strategy that insures against governance uncertainty and show that users who quickly reconnect with prior contacts remain active longer on Bluesky.
}

\section{Introduction}
Large social media platforms, such as Twitter, now renamed to X, mediate much of today’s professional and public communication. As such, these platforms' policies shape everything from general political discourse \autocite{bailExposureOpposingViews2018,cinelliEchoChamberEffect2021,mccabePostJanuary6thDeplatforming2024,quelleBlueskyNetworkTopology2025} to academic discussions \autocite{gargPoliticalExpressionAcademics2025,mohammadiAcademicInformationTwitter2018}. Yet little is known about what drives users from one platform to another, as large-scale, coordinated switches are rare. This rarity stems from a defining feature of social media, strong network effects, which create high switching costs, as the value of participation increases with the number of users, making collective exit difficult once a platform is established \autocite{katzNetworkExternalitiesCompetition1985}. These dynamics often result in winner-take-all outcomes, as each new user further enhances the platform’s value. These network externalities reinforce the dominance of the incumbent platform \autocite{galHiddenCostsFree2016,farrellChapter31Coordination2007} even more when departure means severing ties to established audiences and collaborators. Users dissatisfied with governance, features, or content policies may therefore remain, fearing the loss of both reach and relationships—a pattern consistent with evidence of negative externalities extending even to non-users \autocite{bursztynWhenProductMarkets2023}. The forces that bind users to the platforms are largely unobservable in the day-to-day steady state as small fluctuations do not meaningfully perturb behavior. Identifying these forces requires prolonged disruption large enough to move the system away from its equilibrium.

Many users instead engage in multi-homing, which means they maintain partial presence across multiple platforms without fully abandoning incumbent platforms \autocite{rochetPlatformCompetitionTwoSided2003,JudgementBytedanceLtd2024}. Despite extensive theory on coordination and critical-mass dynamics\autocite{granovetterThresholdModelsCollective1978,oliverTheoryCriticalMass1985,centolaExperimentalEvidenceTipping2018}, opportunities to observe large enough perturbations to identify the forces sustaining dominant platforms are rare\autocite{delreTargetingTimingPromotional2007,golderGrowingGrowingGone2006}. Most established services have maintained stable user bases, and even during moments of discontent, such as the 2015 Reddit controversy\autocite{newellUserMigrationOnline2016} or the early post-acquisition turmoil at Twitter\autocite{ngJournalistsExodusNavigating2025}, any migration was partial and temporary \autocite{jeongDescriptorTemporalMultinetwork2024}. Alternative platforms rarely achieved sufficient density to sustain long-term engagement. The recent Twitter-to-Bluesky migration, however, represents an unusually broad and sustained shift, offering a rare opportunity to study collective behavior as large numbers of users partially exit an incumbent platform.

We build on prior work that treats migration as a form of collective behavior driven by social influence \autocite{jeong2024exploring, radivojevicUserMigrationTwitter2025, bittermann2025social, fieslerMovingLandsOnline2020,cavaDriversSocialInfluence2023,heFlockingMastodonTracking2023}. Yet earlier studies have been constrained by small samples\autocite{jeong2024exploring,fieslerMovingLandsOnline2020}, selection bias toward self-declared transitioned users\autocite{jeong2024exploring,cavaDriversSocialInfluence2023,fieslerMovingLandsOnline2020,heFlockingMastodonTracking2023}, and limited data on pre-migration behavior and network structures\autocite{radivojevic2024reputation}. Our study overcomes these limitations by analyzing the population of 276{,}431 academics active on Twitter between 2023 and early 2025, a community for whom online visibility and professional identity are tightly linked, which facilitates linking users across platforms \autocite{mohammadi2018academic,chanTwitterCitations2023,jungerDoesReallyNo2020,holmbergArticlesOpenAccess2020,wangFailedMigrationAcademic2024,zheng2025science}. Linking these records to complete Bluesky account data enables the first large-scale, individual-level analysis of actual migration rather than expressed intent.

Disentangling homophily from social contagion is difficult because similar people select into shared networks and encounter common environments and shocks, which align adoption times and produce clustering that is observationally equivalent to peer influence \autocite{aral2009distinguishing,shaliziHomophilyContagionAre2011,sornette2004,Crane2008}. In this setting adoption can follow simple contagion where each exposed neighbor independently raises risk, or complex contagion where reinforcement from multiple contacts or a threshold fraction of adopting neighbors is required \autocite{andersonInfectiousDiseasesHumans1991,watts2002simple,centola2007complex,granovetterThresholdModelsCollective1978,morris2000contagion}. In threshold models, adoption decisions arise from weighing the benefits of coordination with adopting neighbors against the costs of abandoning the status quo \autocite{granovetterThresholdModelsCollective1978,morris2000contagion,easley2010networks}. The literature often emphasizes complex contagion for costly or identity laden behaviors and links it to tipping dynamics once participation clears a threshold \autocite{centola2007complex,centolaExperimentalEvidenceTipping2018,oliverTheoryCriticalMass1985,guilbeault2021topological}. Distinguishing these mechanisms hinges on timing and local structure, so designs that condition on pre-existing similarity and leverage exogenous variation are needed to separate influence from clustered responses \autocite{aral2009distinguishing,shaliziHomophilyContagionAre2011,cencettiDistinguishingSimpleComplex2023}.

Academics represent an ideal study population for two reasons. First, disentangling social contagion from homophily requires observing a tightly connected community. Academic networks are densely interlinked, making them well-suited for studying contagion dynamics. Second, cross-platform tracking requires users to maintain stable, identifiable personas, a criterion academics satisfy given their reliance on consistent public identities tied to professional affiliations \autocite{jungerDoesReallyNo2020,chanTwitterCitations2023,gargPoliticalExpressionAcademics2025}.

Here, we show that one in five academics adopted Bluesky. Leveraging the large and heterogeneous population of academics, we demonstrate significant differences in migration propensities across disciplines, demographics, and social media usage intensity. Adoption varies systematically with investment in Twitter/X—posting volume, reach, network centrality—while academic prestige and citation metrics add little explanatory power. Political expression systematically predicts transition, with engagement on topics traditionally associated with liberal or progressive positions showing the strongest associations. We test whether peer influence accounts for this heterogeneity with two complementary causal analyses. We exploit Brazil’s court-ordered nationwide suspension of X. The ban severed interactions for Brazilian users and generated exogenous variation in peer exposure, allowing us to isolate peer influence from confounding factors. We complement this natural experiment with a daily dynamic matching design that pairs each user with similar peers who have not yet been exposed, comparing like with like as risk sets evolve. Naive exposure comparisons inflate peer effects by roughly an order of magnitude. Corrected estimates reveal influence that is short-lived and dose-dependent. The risk of adoption approximately doubles within a day of a peer's move and returns toward baseline within a week, with mutual ties producing the strongest effects. 

To clarify the mechanisms underlying these peer effects, we apply a contagion–type discrimination method\autocite{cencettiDistinguishingSimpleComplex2023} and extend a node-level classification framework\autocite{andres2025} to the directed Twitter/X network. We simulate mixed-mechanism cascades that combine simple contagion, complex contagion, spontaneous, and shock-driven adoption, and train a classifier on these synthetic events to label each observed transition. This mechanism-level decomposition shows that simple contagion is the predominant driver of migration, exogenous shocks generate short-lived surges, and complex contagion is comparatively rare. Politically inactive users are significantly more likely to transition during shocks or only after repeated reinforcement, indicating higher adoption thresholds and lower sensitivity to single peer cues than politically active users. On the other hand, highly invested users hedge governance risk by establishing a parallel presence that allows rapid reconstruction of their social graph \autocite{fang2021platform}. This pattern is consistent with multi-homing as insurance. Multi-homing removes the need to coordinate with one's peers on Twitter/X, leading to a simple rather than complex contagion mechanism. By early 2025, only around one in five transitioned users had deleted their X account, compared with just over one in ten non-transitioners, and transitioned users remained substantially more active on Twitter/X (around one in six inactive, versus nearly one in three among non-transitioners). We further find that reconstruction of one’s network on Bluesky during the first interaction with the platform significantly predicts longer tenure and higher engagement, with effects on the order of ten percent or more across key metrics. This indicates that interoperability, making reconstruction of a user's network easier, could increase user engagement and the viability of entrant platforms. 

\section{Results}
We link a longitudinal panel of 276{,}431 academics on Twitter/X (identified in the \citeauthor{gargPoliticalExpressionAcademics2025}\autocite{gargPoliticalExpressionAcademics2025} dataset and enriched with OpenAlex bibliometrics) to their new profiles among a complete census of 16.7 million Bluesky accounts collected via the public API (cut-off December~10, 2024). Cross-platform identities are linked with a two-stage procedure—MinHash/LSH blocking followed by a DistilBERT-based entity matching model\autocite{li2020deep}—which attains a macro-F1 of 0.96 (Table~S1). This entity-resolution approach proves necessary because only 53\% of matched academics use identical handles across platforms, meaning naive username matching would miss nearly half of all transitions. We define a transition as the creation of a Bluesky account by an academic in our Twitter/X panel. We focus on transitions rather than profile deletions because account creation captures the key first step in migration and allows us to study partial exit and multi-homing, which are central to our argument. Transitions, therefore, provide a direct measure of entry into the alternative platform and are strongly associated with subsequent Twitter/X disengagement. By early 2025, transitioned users were 62\% more likely to have deleted their Twitter account (18.6\% vs. 11.5\% of non-transitioners) and 43\% less likely to be inactive (17.9\% vs. 31.4\%).  The Results section proceeds in four steps. (i) We document heterogeneity in transition rates, (ii) quantify peer influence above and beyond homophily with matched designs and a quasi-experimental design leveraging an external shock to the network, (iii) disentangle simple, complex, shock-driven, and spontaneous contagion mechanisms, and (iv) show that preserving one's followee network on Bluesky predicts sustained engagement.

\subsection{Heterogeneity in transition rates}

\begin{figure}[!ht]
    \centering
    \includegraphics[width=\linewidth]{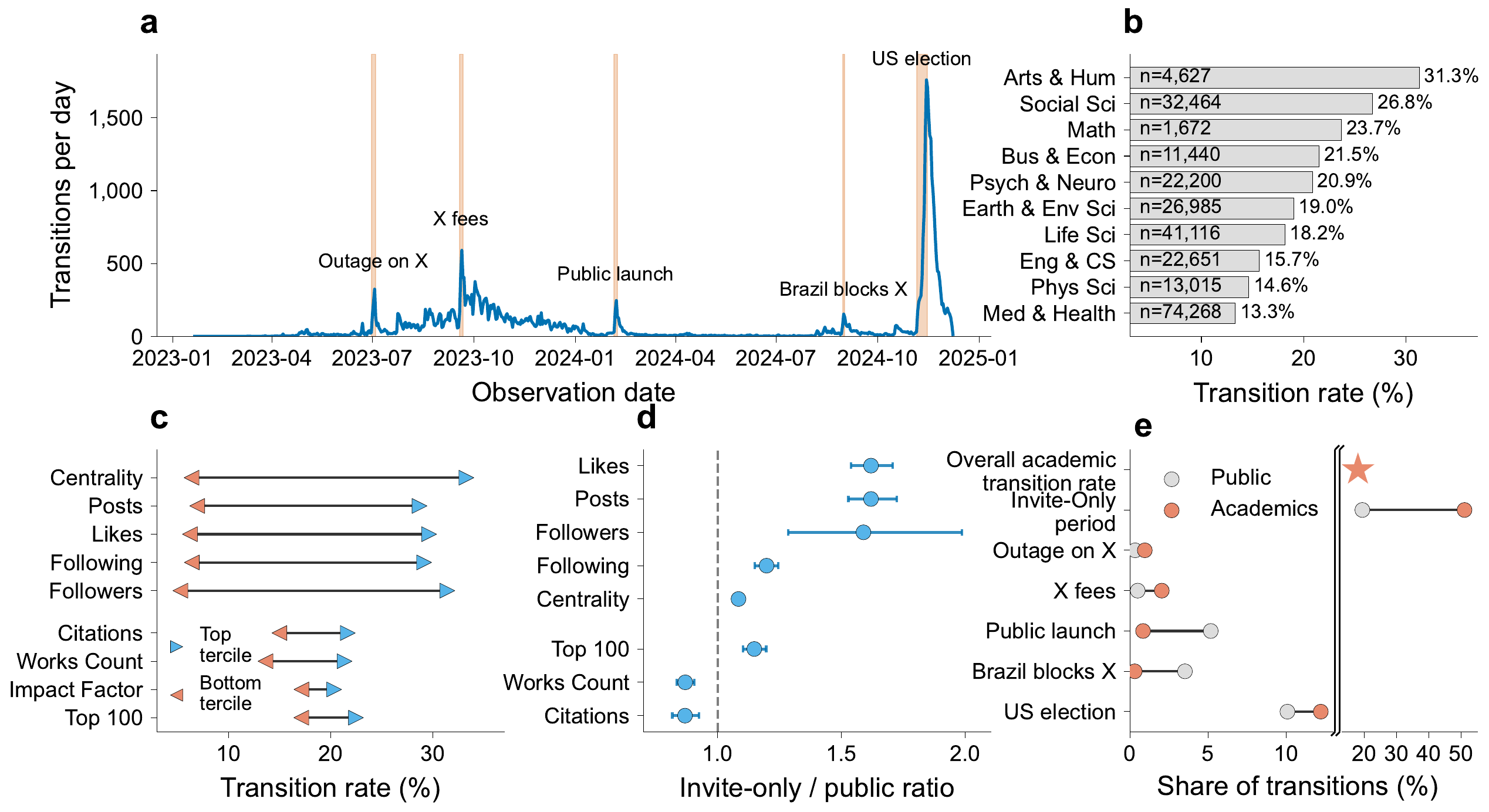}
    \caption{\textbf{Descriptive patterns of the academic migration from Twitter/X to Bluesky.} \textbf{a} Daily number of academic transitions over time, highlighting five major shock periods that drove concentrated migration waves, with the November 2024 US election producing the largest spike (see Table~S10 for detailed discussion). \textbf{b} Transition rates by academic discipline, revealing substantial variation from 13.3\% in Medicine \& Health to 31.3\% in Arts \& Humanities, with sample sizes shown for each field. \textbf{c} Transition rates by user characteristics, comparing top tercile (blue arrows, right) and bottom tercile (red arrows, left) users across academic and social media metrics. Twitter-specific indicators show stronger associations with transition behavior than traditional academic metrics. PageRank is used to compute the centrality of academics in the Twitter network. \textbf{d} Ratios of user characteristics during invite-only versus public release periods, showing early adopters had higher Twitter engagement but lower academic productivity (99\% bootstrapped CIs, 1,000 replicates). \textbf{e} Share of total sign-ups on Bluesky during major shock events, comparing academics (triangles) versus the general public (circles). The US presidential election (red star) generated the largest migration surge }
    \label{fig:descriptives}
\end{figure}

In total, 18\% of academics in our dataset transitioned.
Figure~\ref{fig:descriptives}a displays the daily number of academics who transitioned to Bluesky during the observation period. We highlight shock periods (see Table~S10) in red. The time series shows a persistent background level of transitions with noticeable spikes around several external events.
The overall transition rate masks significant heterogeneities. Among users who have interacted with Twitter in any way since 2024, the transition rate was over 50\% higher, with 27.15\% transitioning. Variation is also visible across academic disciplines as shown in Figure \ref{fig:descriptives}b. Academics in fields such as Medicine and Health (13.3\%), Physical Sciences (14.6\%), and Engineering and Computer Science (15.7\%) saw relatively low transition rates. In contrast, the Social Sciences (26.8\%) and Arts \& Humanities (31.3\%) exhibited much higher adoption of Bluesky.

Transition behavior also correlates with academic credentials and Twitter metrics. Figure \ref{fig:descriptives}c reports transition rates among users in the top and bottom terciles of various academic and social media indicators. While traditional academic metrics show limited predictive power, we find that Twitter-specific indicators (such as PageRank centrality, follower count, and posting frequency) are strongly associated with transition behavior. This suggests that a user’s embeddedness in the Twitter ecosystem is more important than academic standing in predicting early platform migration, mirroring previous findings \autocite{radivojevicUserMigrationTwitter2025}. 

Until February 6, 2024, early adopters of Bluesky required an invite code to join the platform, making existing social networks particularly crucial during this initial phase of platform growth. Figure \ref{fig:descriptives}d highlights this by plotting the difference in the means of key user characteristics between invite-only and public-release cohorts. Users who joined during the invite-only period were significantly more active and central on Twitter, having 8\% higher PageRank centrality, 20\% more followings, 59\% more followers, and 62\% more posts, suggesting that Twitter connections were a pathway to securing early invites. Early adopters were 14.9\% more likely to be affiliated with top-100 universities. However, they showed lower academic productivity with 13.1\% fewer works and 13.2\% fewer citations than later adopters. This indicates Bluesky's initial adoption among academics was driven more by social media engagement than traditional academic credentials.

While we focus on academic users, we can compare the relative proportions of sign-ups per period to those of the general population (Figure~\ref{fig:descriptives}e). Academics were significantly more likely to sign up during the invite-only period, with around half of all academic transitions occurring before the policy change, whereas only about one in five members of the general public signed up during that time. Because a larger share of academics had already transitioned by the time Bluesky opened to the public, later shocks (Public launch, Brazil's X ban) drew from a smaller remaining pool of users who had not yet transitioned, resulting in lower shares of academic transitions during these events. By contrast, the US election—occurring well after the public launch—still produced a slightly higher share of academic transitions, suggesting this event uniquely mobilized academics who had previously resisted switching.

\subsection{Political expression}
We test whether public engagement with political topics predicts migration, controlling for demographic and behavioral covariates shown in Figure~\ref{fig:descriptives}. While we cannot directly observe users’ political orientation, prior work has shown that academic Twitter tends to lean left-of-center \autocite{alabresePoliticizedScientists2024}, and that political discourse is often a key part of scholarly identity-building online \autocite{gargPoliticalExpressionAcademics2025}. Rather than inferring ideology, we focus on expression—that is, the extent to which users choose to engage with political, economic, and social issues in their tweets. Using a set of 14 pre-selected topics (see Methods), we code whether users discussed socio-political subjects. 

\begin{figure}[!ht]
    \centering
    \includegraphics[width=\linewidth]{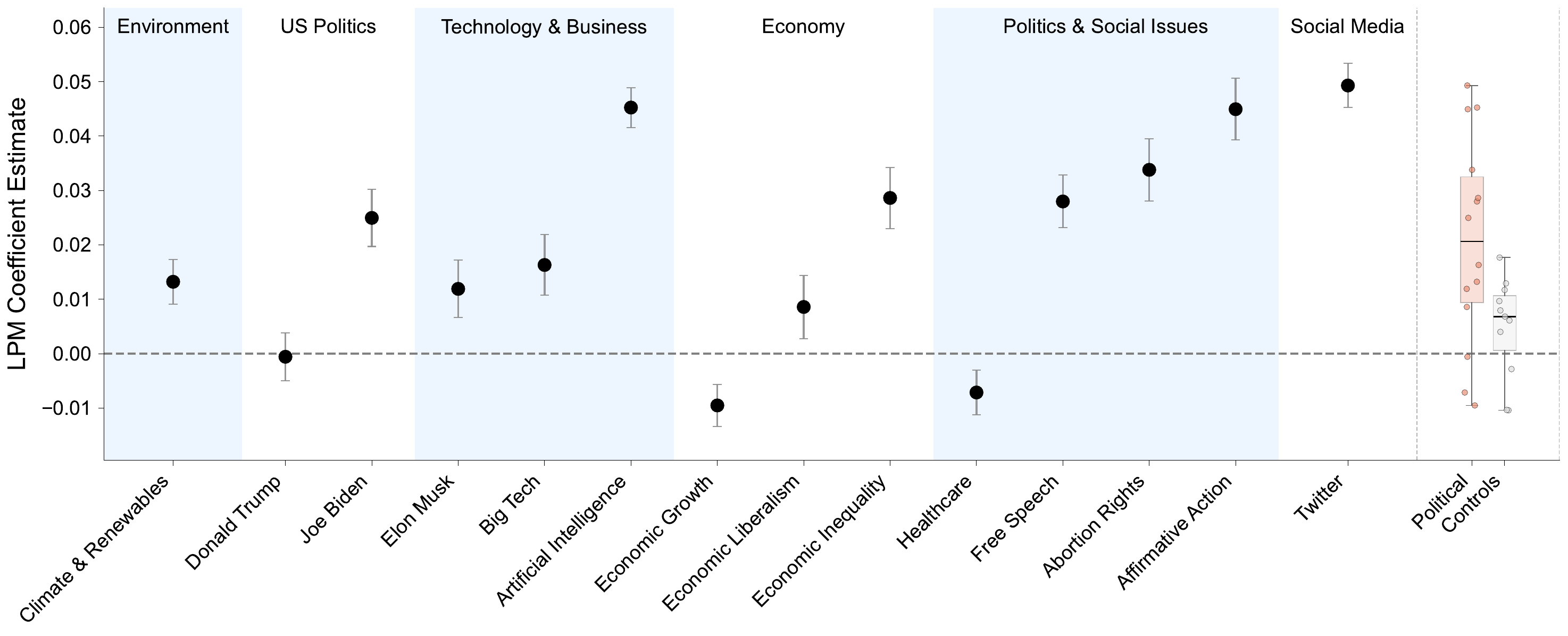}
    \caption{\textbf{Political expression and transition to Bluesky.} Linear Probability Model (LPM) coefficients showing associations between expressing opinions on political topics and Twitter-to-Bluesky transition probability ($N = 206{,}898$; $\text{df} = 206{,}819$). Points represent joint-model estimates that include all topic indicators alongside the same baseline controls. Boxplots on the right summarize the distribution of coefficients for political topics (red) versus non-political controls (gray). Error bars indicate 95 confidence intervals.}
    \label{fig:polexpr}
\end{figure}

Figure~\ref{fig:polexpr} reports Linear Probability Model (LPM) estimates of transition probability as a function of topic engagement. The LPM is a regression with a binary outcome—in this case, whether a user transitioned—which allows for straightforward interpretation of coefficient magnitudes as changes in probability. All models include demographic and behavioral covariates (disciplinary field, gender, cumulative Twitter activity, and 2022–23 posting frequency). We also include non-political topics (e.g., cooking, pets, sports) to benchmark political effects against everyday discourse. The full set of results for all control topics is provided in Table S5, and the demographic, country, and field control coefficients are reported in Table S6.

The three topics most strongly associated with transitioning to Bluesky, after controlling for baseline covariates and all other topics, are posts that mention Twitter, affirmative action, and artificial intelligence. The most strongly negative coefficient is for posts referencing economic growth, possibly reflecting relatively more conservative users who fail to adopt \autocite{allgoodEconomicsCourseworkMajoring2012}. This is consistent with findings that Bluesky’s user base is overwhelmingly left-of-center \autocite{quelleBlueskyNetworkTopology2025}.

To distinguish political expression from more general engagement with popular social media topics, we also include non-political controls, such as references to cooking, gardening, travel, or sports, that are designed to be mostly uncorrelated with political topics. While some of these are significantly associated with transition probability, their effect sizes are markedly smaller. Taken together, the results indicate that publicly articulated positions—particularly on artificial intelligence, affirmative action, and Twitter’s governance—are systematically related to early adoption of Bluesky. While descriptive, it highlights political expression as an important pre-treatment dimension of heterogeneity that helps explain observed variation in migration patterns.

\subsection{Estimating network effects with a natural experiment}
\begin{figure}[!ht]
  \centering
  \includegraphics[width=\textwidth]{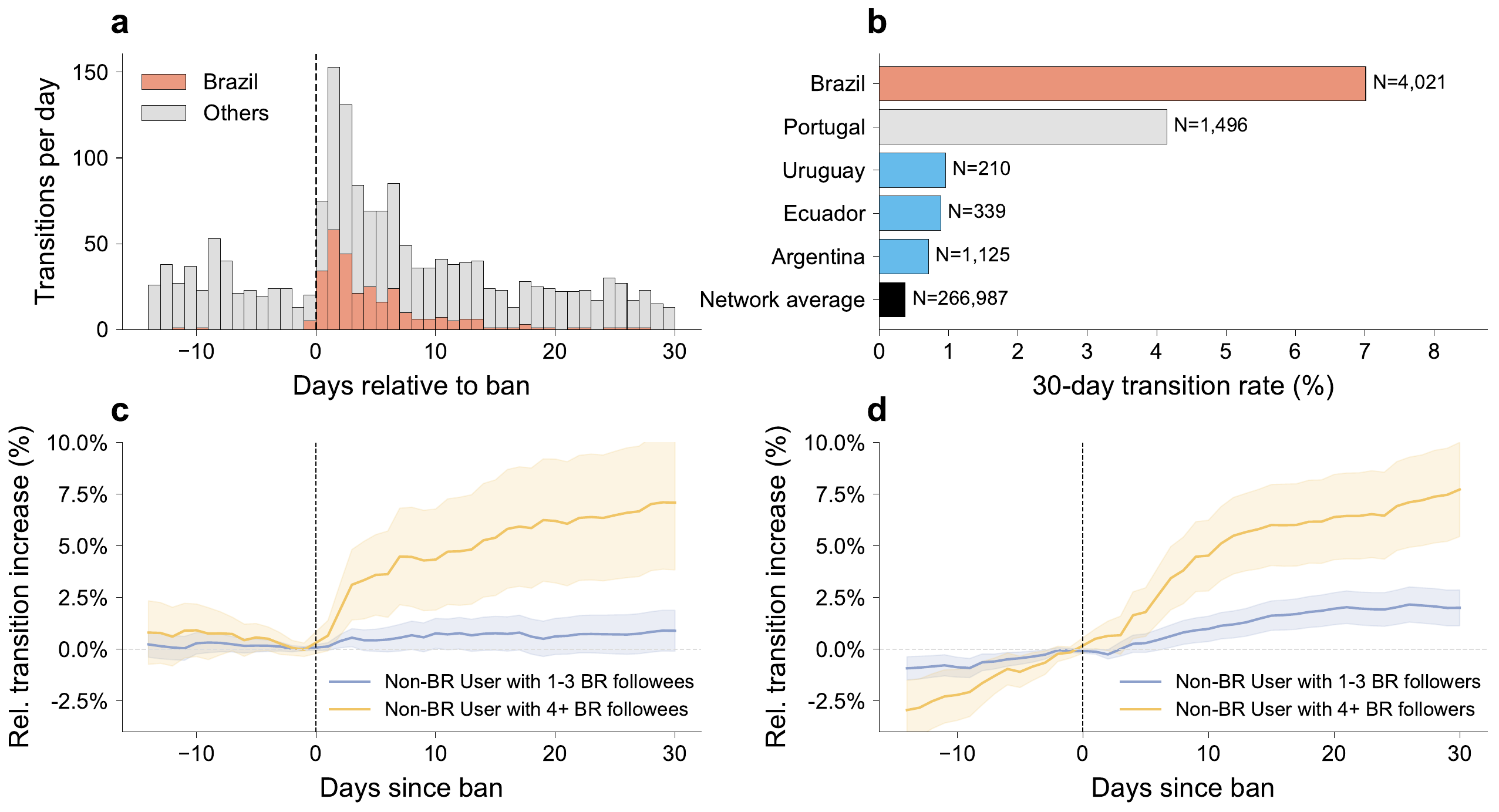}
\caption{\textbf{Brazil's suspension of Twitter/X as a natural experiment.} \textbf{a} Absolute transitions in 24-hour bins from day $-14$ to day~30, split into Brazilian (red) and others (gray). \textbf{b} Country-level 30-day transition rates among sizeable groups (top five shown) plus the network average (black). \textbf{c,d} Event-study estimates showing the relative probability of transitioning around day~0 for non-Brazilian scholars, grouped by number of Brazilian ties. We compare treated users against users with no Brazilian connections ($N = 8{,}645{,}021$ user-days, df = $192{,}508$). Panel \textbf{c} uses Brazilian followees and panel \textbf{d} uses Brazilian followers. Shaded areas denote 95\% confidence intervals. Sample sizes for Followees: 0 ties = $202{,}729$, 1--3 ties = $31{,}592$, 4+ ties = $5{,}596$ and for Followers: 0 ties = $187{,}789$, 1--3 ties = $42{,}618$, 4+ ties = $9{,}510$.}
  \label{fig:brazilshock}
\end{figure}

Observable clustering of adoption among connected academics does not, by itself, establish peer influence. Clustering of adoptions among social ties may be explained by homophily, where similar people react to the same external cues that cause adoption \autocite{aral2009distinguishing,shaliziHomophilyContagionAre2011}. We separate peer influence from homophily through two complementary designs. The first leverages the fact that on August 30, 2024, access to Twitter/X was abruptly suspended in Brazil following a judicial order, producing an externally imposed, country-specific disruption (day 0). The intervention’s timing and geography were entirely exogenous to user behavior and confined to Brazil, creating sharp variation in exposure among non-Brazilian academics depending on their pre-existing ties to Brazilian accounts. This generates a natural experiment in which users are comparable ex ante but differ in whether their networks were directly affected by the ban. In Figure~\ref{fig:brazilshock}a, we plot the absolute number of transitions in our sample in daily bins around the shock. We find that Brazilian departures are near zero before the shock, spike immediately at day 0 (accounting for a plurality of departures at peak), and then ebb over subsequent weeks, with a smaller rise outside Brazil. Figure \ref{fig:brazilshock}b then reports 30-day transition rates by country (top four shown) plus the network average. Brazil is most affected (\emph{N} = 4{,}021; $\approx 7\%$ over the next 30 days), far above the network average ($0.37\%$); among countries with at least 200 academics, Portugal (shared language) as well as the geographically closer Uruguay, Ecuador and Argentina also sit above the mean. With this, we have established that the shock had a large impact on the Brazilian academic users' likelihood to transition and given some indicative evidence that although they are not directly affected, several geographically or culturally close countries also had high transition rates in the days following the ban. However, this is not enough to attribute this to network effects. We therefore exploit this quasi-experiment to estimate the causal network effect (Figure~ \ref{fig:brazilshock}c,d). We compare non-Brazilian users with varying numbers of Brazilian connections (none, 1–3, and $\geq$ 4), separately for ties to information sources (followees) and audiences (followers). Because exposure is determined solely by pre-shock network structure, and the ban itself is external, any post-shock divergence is consistent with contagion through network ties rather than shared predispositions. Consistent with this interpretation, transition behavior is mostly parallel across groups before the ban, particularly for followee-based ties, reinforcing the identification assumption. This suggests that in the absence of the ban, there would likely not have been a large migration of non-Brazilian users connected to Brazilians. After day 0, transition rates rise sharply with the number of Brazilian followers and followees.  The most affected users, i.e., those with more than four Brazilian followees, are more than 7\% more likely to transition than those unexposed to the first-order network effects. This effect is highly concentrated on the first seven days, which account for two-thirds of the overall increase.  On the other hand, small pre-trends among the follower groups suggest that these estimates likely represent upper bounds on the true effect for followers. Taken together, the Brazil suspension offers a rare natural experiment that isolates peer contagion from homophily, owing to its externally imposed timing and geographical scope.

\subsection{Estimating network effects with dynamic propensity score matching}
To broaden our sample and speak more generally about the mechanisms driving these peer effects across the entire sample period, we implement a dynamic matched–sample design using generalized propensity score matching \autocite{aral2009distinguishing,HiranoImbens2004}. Treatment is defined as exposure to a neighbor who transitioned to Bluesky, with timing designs capturing recency of exposure (1–6 days) and dose designs capturing intensity (number of transitions in the past week). We estimate propensity scores from 225 pre-treatment covariates capturing dynamic network saturation, individual attributes, and averaged characteristics of each user's audience and information sources (Table~S7). All covariates are measured at least one week before the treatment window to ensure strict temporal separation. Within propensity-score calipers, we select controls via approximate nearest-neighbor matching on 11 core network variables \autocite{HiranoImbens2004} (see SI Section~4 for diagnostics). This design compares users who are ex ante equally likely to experience exposure, conditional on observables, but differ in whether a connection adopted Bluesky.

\begin{figure}[!ht]
    \centering
    \includegraphics[width=\linewidth]{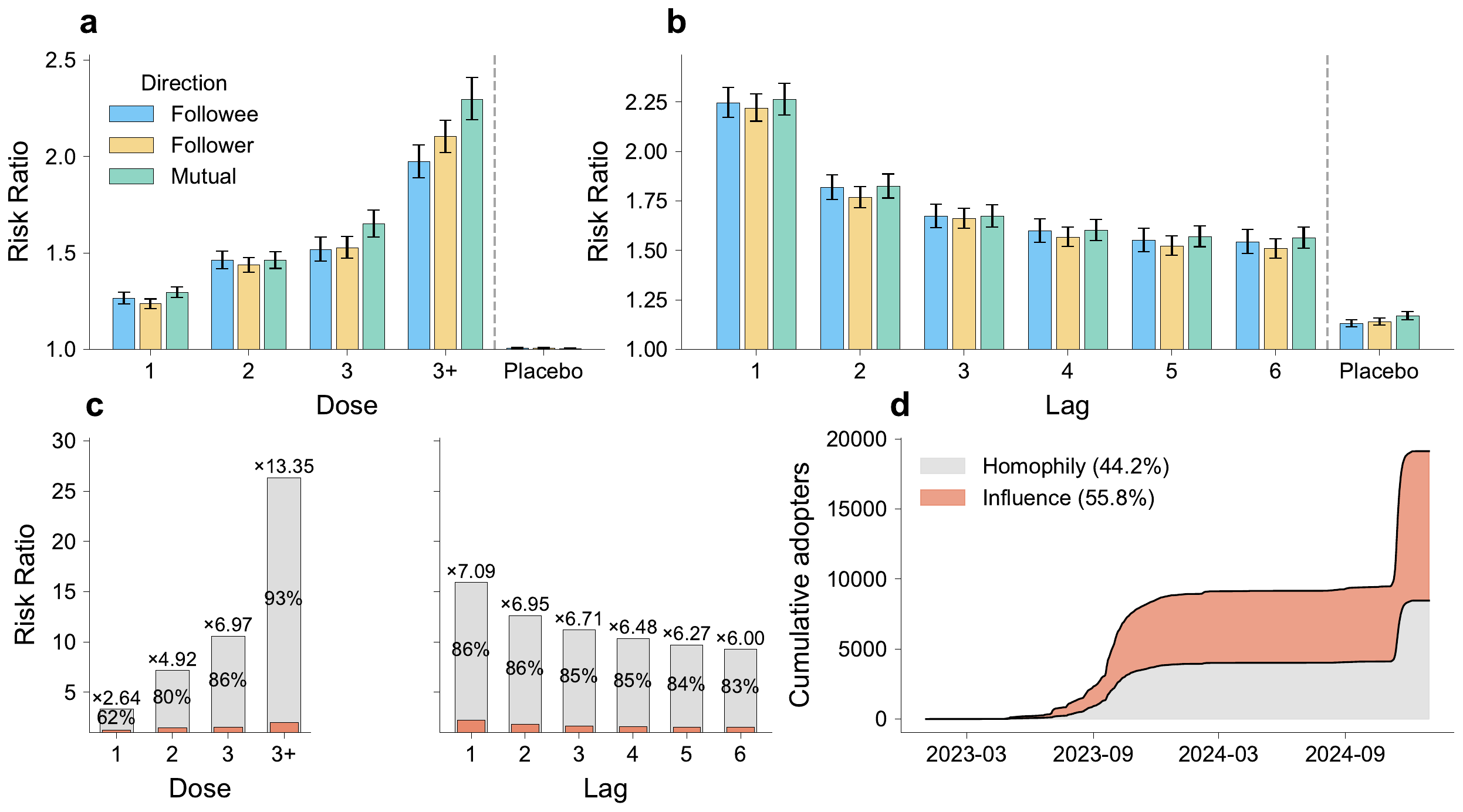}
    \caption{\textbf{Dynamic matching estimates of peer influence.} \textbf{a} Risk ratios by dose (number of connections who left in the prior week), disaggregated by tie direction. Total matched pairs: followees 58.0M, followers 52.1M, mutual 33.7M. \textbf{b} Risk ratios by timing (days since a connection left). Total matched pairs: followees 66.8M, followers 61.0M, mutual 38.9M. \textbf{c} Comparison of naïve and matched causal estimates. \textbf{d} Decomposition of treated adoptions into contagion and homophily. Error bars show 95\% CIs. Cumulative sample sizes per treatment condition are reported in Tables S7–S8.}
    \label{fig:fig4_causal}
\end{figure}

Figure~\ref{fig:fig4_causal} reports causal estimates from the dynamic matching design. We measure peer influence using risk ratios (RR), defined as the transition probability of exposed users relative to matched unexposed controls. Figure~\ref{fig:fig4_causal}a shows that transition risk increases with the number of connections who left in the prior week. Mutual ties consistently produce the largest effects, with followees exceeding followers at most doses. Placebo doses, created by randomly shuffling exposures across the network, are indistinguishable from zero.

Figure~\ref{fig:fig4_causal}b reports timing effects that decay rapidly with lag. Exposure to a connection that left yesterday more than doubles the daily transition risk relative to matched controls, with smaller effects for older events. Directional differences mirror Figure~\ref{fig:fig4_causal}a. While significant, the effects of placebo treatments are negligible compared to causal treatments. 

Figure~\ref{fig:fig4_causal}c compares naïve estimates of the transitioning probability to our matched causal estimates for followee exposure. Naïve risk ratios substantially overstate influence, with 62\% of the apparent effect at dose 1 and up to 93\% at dose $3{+}$ attributable to homophily rather than contagion. Similarly, for timing-based treatments, only 15\% of observed clustering is attributable to contagion. 
Figure~\ref{fig:fig4_causal}d attributes treated adoptions over time for the $d{=}1$ design, indicating that 55.8\% of adoptions among treated users are explained by social influence and 44.2\% by homophily. We decompose the adoption into homophily and contagion by multiplying the daily control-to-treated adopter ratio from our matched sample by the daily population of treated adopters.

\subsection{Contagion mechanisms}
\label{methods:contagion}
Contagion research distinguishes three core processes\autocite{andersonInfectiousDiseasesHumans1991, morris2000contagion,watts2002simple,centola2007complex}. Simple contagion follows epidemiological logic, where each infected neighbor independently increases adoption probability. Complex contagion requires social reinforcement through multiple exposures or a critical local fraction of contacts who have already adopted\autocite{granovetter1978threshold, cencettiDistinguishingSimpleComplex2023}. Spontaneous adoption captures idiosyncratic moves unrelated to peer behavior. Because our data show five abrupt surges coinciding with external events, we extend the framework of \citeauthor{andres2025}\autocite{andres2025} to include exogenous shocks as a fourth mechanism following models of shock-driven dynamics in social systems\autocite{sornette2004,Crane2008} (see Methods). Following prior work, we do not explicitly model homophily in our node-level classifier \autocite{andres2025, cencettiDistinguishingSimpleComplex2023}. This simplification is justified because the homophily-attributable share among treated adopters is stable (44\%; weekly mean 46\%, median 48\%, with 75\% of weeks within $\pm$10 p.p.) and adopter composition is similar across low- vs. high-homophily weeks. We calibrate parameters to empirical adoption patterns, enabling classifiers trained on synthetic cascades to decompose the entire migration wave at individual resolution. We train a machine learning classifier on synthetic cascades to identify which mechanism drives each individual's transition.

\begin{figure}[H]
    \centering
    \includegraphics[width=1\linewidth]{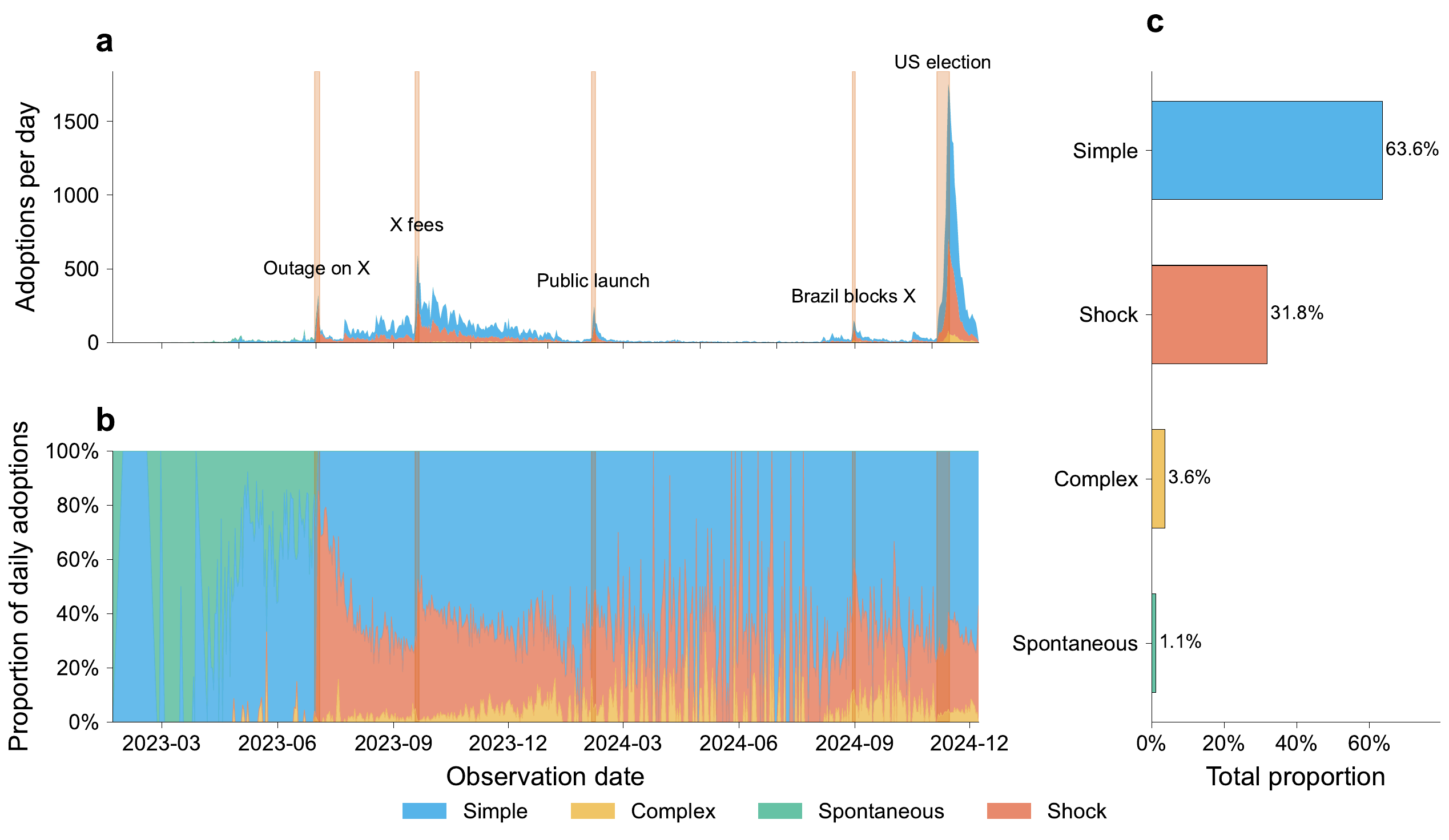}
    \caption{\textbf{Contagion mechanism decomposition based on a machine learning classification.} \textbf{a} Daily number of academic transitions by contagion mechanism, showing five major shock periods where shock-driven transitions (red) create the large spikes. \textbf{b} Daily proportion of transitions by contagion mechanism. \textbf{c} Overall distribution of transition mechanisms across the study period ($N = 49{,}510$ users: Simple 63.6\%, Shock 31.8\%, Complex 3.6\%, Spontaneous 1.1\%).}
    \label{fig:transition_mechanism}
\end{figure}

Simple contagion emerges as the dominant mechanism, accounting for two-thirds of all transitions. The relative importance of mechanisms varies dramatically over time. In the earliest stages, spontaneous transitions represent a disproportionately high share, as first movers cannot be influenced when few connections have yet transitioned. During baseline periods, simple contagion consistently drives the majority of transitions. Shock events create temporary surges, with shock-driven adoption accounting for up to 60\% of daily transitions. The November 2024 US presidential election produced the largest spike. Complex contagion represents the smallest share at 3.6\% of all transitions but notably increases after Bluesky's public release in February 2024. The \citeauthor{cencettiDistinguishingSimpleComplex2023}\autocite{cencettiDistinguishingSimpleComplex2023} degree-order test independently confirms this pattern ($\rho = -0.21$, $p < 0.001$, Figure~S2).

\begin{table}[htbp]
  \centering
  \resizebox{\textwidth}{!}{%
  \begin{tabular}{lcccccc}
  \toprule
   & Simple & Shock & Complex & Spontaneous & $\chi^2$& $V$ \\
  \midrule
  \textbf{N (\%)} & 31,477 (63.6\%) & 15,720 (31.8\%) & 1,785 (3.6\%) & 528 (1.1\%) & & \\
  \addlinespace
  \textbf{Twitter Investment} & & & & & & \\
  \quad Bottom quartile & 47.6\% & 41.6\% & 9.8\% & 1.1\% & \multirow{2}{*}{2084.9***} & \multirow{2}{*}{.290} \\
  \quad Top quartile & 71.9\% & 25.3\% & 0.8\% & 2.0\% & & \\
  \addlinespace
  \textbf{Political Engagement} & & & & & & \\
  \quad Bottom quartile & 49.8\% & 41.4\% & 7.8\% & 1.0\% & \multirow{2}{*}{1555.2***} & \multirow{2}{*}{.250} \\
  \quad Top quartile & 71.9\% & 24.9\% & 1.5\% & 1.7\% & & \\
  \addlinespace
  \textbf{Activity} & & & & & & \\
  \quad Bottom quartile & 59.8\% & 33.5\% & 5.2\% & 1.4\% & \multirow{2}{*}{193.8***} & \multirow{2}{*}{.098} \\
  \quad Top quartile & 66.6\% & 29.7\% & 2.1\% & 1.6\% & & \\
  \addlinespace
  \textbf{Academic Prestige} & & & & & & \\
  \quad Bottom quartile & 59.4\% & 33.8\% & 5.5\% & 1.3\% & \multirow{2}{*}{138.3***} & \multirow{2}{*}{.075} \\
  \quad Top quartile & 65.0\% & 31.0\% & 3.1\% & 0.9\% & & \\
  \addlinespace
  \textbf{Deletion} & & & & & & \\
  \quad Deleted & 64.7\% & 30.8\% & 3.6\% & 0.9\% & \multirow{2}{*}{8.0*} & \multirow{2}{*}{.013} \\
  \quad Not Deleted & 63.3\% & 32.0\% & 3.6\% & 1.1\% & & \\
  \bottomrule
  \multicolumn{7}{l}{\footnotesize *$p<.05$, ***$p<.001$. $V$ = Cramer's V. Rows sum to 100\% across mechanisms.}
  \end{tabular}%
  }
  \caption{\textbf{Transition mechanism profiles by user characteristics.} Each row shows the distribution of contagion mechanisms within that subgroup. \textit{Twitter Investment} = average of percentile ranks for follower count, post count, and network centrality. Bottom quartile medians: 198 followers, 190 posts. Top quartile medians: 3,851 followers, 8,070 posts. \textit{Political Engagement} = total number of tweets on political topics. Bottom quartile median: 2 tweets. Top quartile median: 156 tweets. \textit{Activity} = change in post count between 2022 and 2025 among users whose accounts were not deleted. Bottom quartile: $\leq$3 posts. Top quartile: $\geq$365 posts. \textit{Academic Prestige} = average of percentile ranks for publication count and citation count. Bottom quartile medians: 5 works, 16 citations. Top quartile medians: 125 works, 2,955 citations. \textit{Deletion} = Twitter account unavailable in 2025.}
  \label{tab:mechanism_profiles}
  \end{table}

The dominance of simple contagion in aggregate masks important heterogeneity in who responds to which mechanism (Tab. \ref{tab:mechanism_profiles}). Transition mechanisms vary strongly with Twitter investment and political engagement. Users in the top quartile of Twitter investment or political engagement follow simple contagion roughly 70\% of the time, compared to less than 50\% for bottom-quartile users, who instead rely more heavily on shocks (42\% vs.\ 25\%) and complex contagion (10\% vs.\ 1\%). Twitter investment shows the strongest association (Cramer's V = 0.29), followed by political engagement (V = 0.25). Activity, academic prestige, and deletion status show negligible associations with mechanism type (V = 0.10, 0.08, and 0.01, respectively). Spontaneous adoption, though rare overall (1.1\%), is most common among top-quartile Twitter investment users (2.0\%).

\subsection{Network effects on Bluesky}
Our analysis thus far has focused on the factors that drive academics to leave Twitter/X for Bluesky. We've established that network effects, mainly dominated by simple contagion, shape who transitions and when. However, successful platform migrations require sustained engagement with the destination platform. If network effects drive transitions but do not predict engagement, this would suggest different mechanisms operate pre- and post-migration. Alternatively, if network preservation is crucial for sustained engagement, successful migrations may require coordinated reconstruction of social connections. We therefore examine whether academics who successfully preserve their Twitter/X networks on Bluesky exhibit higher levels of sustained engagement on the new platform.

\begin{table}[!h]
    \centering
\begin{tabular}{lccccc}
\toprule
Day 1 Reconnection & Active Days & Duration & Posts & Likes & Follows \\
\small{(excl. day 1 actions)} & & & & & \\
\midrule
Day 1 Reconnection & 12.6$^{***}$ & 16.0$^{***}$ & 9.6$^{***}$ & 14.4$^{***}$ & 25.3$^{***}$ \\
Std. Error & (1.9) & (3.2) & (2.0) & (1.4) & (3.1) \\
$p$-value & 0.000 & 0.000 & 0.000 & 0.000 & 0.000 \\
$R^2$ & 0.387 & 0.520 & 0.295 & 0.265 & 0.179 \\
\bottomrule
\multicolumn{6}{l}{\textit{Note:} $^{***}p<0.01$. $N = 37{,}022$; $\text{df} = 36{,}997$.}\\
\multicolumn{6}{l}{Effects show \% increase from 10pp reconnection increase.}
\end{tabular}
\caption{Effect of day-one network reconnection on subsequent Bluesky engagement. Coefficients from OLS regressions of log-transformed outcomes on reconnection rate, with join-month fixed effects. Standard errors clustered by cohort. Effects show percentage increase from 10 percentage point increase in reconnection rate.}
\label{tab:reconnection}
\end{table}

Table \ref{tab:reconnection} reports the relationship between early network reconnection and subsequent platform engagement. We regress log-transformed engagement outcomes—excluding the first 24 hours to avoid mechanical correlation—on the share of available Twitter ties that users followed within their first day on Bluesky. All specifications include join-month fixed effects with standard errors clustered by cohort. Each 10 percentage point increase in day-one reconnection is associated with 12.6\% more active days, 16.0\% longer platform tenure, 9.6\% more posts, 14.4\% more likes, and 25.3\% more follows (all with $p$ < $0.01$). The particularly strong effect on following behavior suggests that early reconnections enable users to reconstruct their social graphs, sustaining engagement across all dimensions.

\section{Discussion}
Platform migrations pose a theoretical puzzle. Network effects should lock users into incumbent platforms because the value of participation rises with each additional user. Yet we document a sustained mass migration of academics from Twitter/X to Bluesky. We show that multi-homing transforms migration from a high-stakes collective action problem into a low-cost individual decision. When users can maintain presence on both platforms, adoption spreads through simple contagion rather than the complex contagion that theory predicts for costly behaviors. The same network forces that drive departure also sustain engagement at the destination. Our analysis draws on a unique dataset of 276{,}431 academics tracked on Twitter/X and Bluesky throughout a period of intense upheaval in social-media governance. This turbulent backdrop reshaped both the costs and benefits of switching platforms and gives us a rare, high-resolution view of how peer influence propagates through follow networks, how individual academics decide when to jump, and what keeps them active once they arrive. 

We find that 18\% of academics transitioned to Bluesky during our observation period, though this average masks substantial variation. Political expression predicts migration, with engagement on topics like abortion rights and affirmative action associated with higher transition probabilities. More surprising is that heavy Twitter users are more likely to transition than light users. Conventional wisdom holds that sunk costs should anchor power users to incumbent platforms \autocite{farrellChapter31Coordination2007,rietveldPlatformCompetitionSystematic2021}. Yet these users are not abandoning Twitter for Bluesky. They are hedging across both, maintaining parallel presence while insuring against platform risk. By early 2025 only 18.6\% of transitioned users had deleted their Twitter accounts and the majority maintained activity on both platforms. Heavy users have the most to lose from governance instability and therefore the strongest motivation to hedge.

Observable clustering of adoptions among connected users does not by itself establish peer influence. Similar people form ties and respond to the same external cues, producing patterns that look like contagion but reflect shared predispositions. We separate genuine influence from homophily through two complementary designs. Brazil's court-ordered suspension of Twitter created exogenous variation in network exposure, and non-Brazilian academics with Brazilian connections became significantly more likely to transition after the ban. We complement this natural experiment with daily dynamic matching that pairs each user with similar peers who differ only in whether their connections have adopted. Naive exposure comparisons inflate peer effects by roughly an order of magnitude. Corrected estimates reveal influence that is genuine but short-lived and dose-dependent. The risk of adoption approximately doubles within a day of a peer's move and returns toward baseline within a week, with mutual ties producing the strongest effects.

The literature on complex contagion predicts that costly or identity-laden behaviors require reinforcement from multiple peers before adoption occurs  \autocite{centola2007complex,granovetterThresholdModelsCollective1978}. An individual facing high switching costs should not adopt after a single peer's move but instead wait until a critical fraction of their network has already transitioned. Platform migration fits this profile. Users must rebuild their audience, learn new interfaces, and risk losing accumulated social capital. Our central finding contradicts this expectation. Simple contagion accounts for two-thirds of all transitions we observe, with each peer departure independently raising the probability of adoption without requiring coordinated thresholds. The resolution lies in switching costs. Multi-homing makes trial nearly costless, and each peer's move provides independent information about Bluesky's viability without requiring social reinforcement.

Transition mechanisms vary systematically with user characteristics. Heavily invested users with large followings, high posting volume, and active political voice are more likely to experience simple contagion, while peripheral and politically disengaged users are more likely to require either repeated social reinforcement from multiple peers or external shocks to overcome inertia. We interpret this heterogeneity as reflecting variation in the push and pull forces users experience, not variation in their ability to multi-home. The option to maintain presence on both platforms lowered switching costs equally for everyone. Heavily invested users are more attuned to platform governance concerns that push them away from Twitter, and they have accumulated more social capital worth protecting, giving them stronger incentives to hedge against platform risk. Our results suggest that, for them, a single peer's signal suffices to tip the balance, producing simple contagion. Peripheral and politically disengaged users face the same low switching costs but experience weaker push and pull. They are less attuned to governance issues and have less to lose from platform instability. We find that transition mechanisms are unrelated to post-transition outcomes. Activity changes and account deletion show negligible associations with mechanism type because these reflect a separate decision. Once users have transitioned, whether to stay active on Twitter or delete their account depends on their experience with Bluesky rather than what triggered their initial move. The mechanism that brings users to a platform does not determine whether they stay.

The network forces that catalyze departure from Twitter also anchor users on Bluesky. Users who rebuild their followee network on day one stay longer, post more, and engage more across every metric we measure. The particularly strong effect on following behavior suggests that early reconnection triggers a virtuous cycle of network reconstruction. This finding speaks directly to ongoing debates about platform interoperability. If social capital can be reconstructed rather than forfeited, network effects need not entrench incumbents as strongly as traditional models suggest.

These results carry implications for platform governance and competition policy. Data portability and network interoperability are often proposed as interventions to reduce platform lock-in  \autocite{rochetPlatformCompetitionTwoSided2003,katzSystemsCompetitionNetwork1994,rietveldPlatformCompetitionSystematic2021}, and our findings suggest these interventions may be more powerful than previously understood. The social reinforcement dynamics thought to entrench incumbents dissolve when users can hedge across platforms. Lowering the cost of reconstructing one's network on a new platform could substantially reduce barriers to competitive entry.

Our analysis has limitations. We study academics, a population that is highly networked and maintains stable public identities across platforms, so the strength and form of peer influence may differ for other user populations. We observe ties and behaviors only on Twitter and Bluesky. Unobserved interactions on other platforms, in offline settings, or through private channels likely attenuate our contagion estimates, while residual homophily that our matching strategy does not fully absorb would inflate them. The migration we study unfolded during a period of unusually salient governance shocks, which may limit extrapolation to more gradual or less politicized transitions while simultaneously providing the strong perturbations that make causal identification possible.

Our results suggest that platform migration spreads like an epidemic when switching costs fall low enough to permit trial. Multi-homing transforms what should be a coordination problem into a series of independent decisions, and the network ties that push users to leave an incumbent platform pull them toward sustained engagement on the challenger. These dynamics suggest that platform competition may be more contestable than network effects theory implies, provided users can port their social capital across platforms.

\section{Methods}
\subsection{Data}
We analyze 276{,}431 scholars on Twitter, using a dataset compiled by \citeauthor{gargPoliticalExpressionAcademics2025}\autocite{gargPoliticalExpressionAcademics2025} extending the 1st version of the dataset in \citeauthor{mongeonOpenDataSet2023}\autocite{mongeonOpenDataSet2023}. The dataset mainly captures established, full-time academics who publish regularly and are embedded in professional networks (see Table~S3). Focusing on established academics is advantageous for our purposes, since they are the scholars most likely to use social media strategically for visibility, networking, and debate, and thus their migration decisions provide a particularly meaningful lens on broader platform dynamics. For each scholar, we have their profile name and biography, directed follower network (i.e., followers and followees), and their complete tweet archive, available via the Twitter API (queried in January 2023), which dates back to 2007. We obtained publication records, institutional affiliations, and disciplinary classifications through linkage with OpenAlex, a freely available dataset containing more than 200 million works matched to authors and their respective institutions, created to replace the discontinued Microsoft Academic Graph \autocite{priem2022openalex}.

\subsection{Cross-platform account identification}
To identify which scholars migrated to Bluesky, we link the full list of Twitter accounts against all known Bluesky users as of December 2024.  This cross-platform account linkage poses a computational challenge, potentially requiring trillions of comparisons between Twitter accounts and 16.7M Bluesky users\autocite{shuUserIdentityLinkage2017}. For the Bluesky users who joined prior to the cutoff date of the data collection, we have information on profile name, description, posts and all follow relations. Given the open nature of the Bluesky platform, this information is easily accessible to researchers via the official API endpoint. 

To establish a reliable ground truth dataset for cross-platform academic user identification, we randomly sampled 1,500 Twitter profiles and manually searched for corresponding matches on Bluesky. Graduate degree holders familiar with both platforms performed all data labeling. To avoid sampling bias, we did not pre-filter accounts based on heuristics, which would have overrepresented ``easy'' matches in our dataset. We supplemented this with a true-negative dataset by generating challenging potential matches for confirmed true-positive cases, resulting in 267 true-positive matches and 801 true-negative matches. 

The fundamental challenge in cross-platform account linking lies in its computational complexity --- an $O(n^2)$ problem requiring about 5 trillion comparisons between nearly 300{,}000 Twitter accounts and 16.7M Bluesky accounts. This scale makes even simple string comparison measures computationally infeasible. To address this, we applied three complementary indices throughout \autocite{papadakisBlockingFilteringTechniques2020}. We complemented two lexical indices based on profile names and handles with one semantic index based on the profile name and description. Using them jointly balances fast character-level matching with meaning-aware retrieval.

Lexical matching approximates Jaccard similarity with MinHash \autocite{eric_zhu_2024_11462182}. We pre-cleaned input strings by normalizing Unicode and removing accents, dropping academic titles (``Dr'', ``Prof'', ``PhD''), and stripping extraneous content. Further, we segment display names into character bigrams and trigrams, whereas handles use bigrams only. From these n-grams, we computed 64 MinHash signatures and indexed them via Locality Sensitive Hashing with a threshold of 0.5. We subsequently re-ranked the candidate set (up to 10,000 profiles) using a token sort ratio \autocite{max_bachmann_2025_17500947}.

To extend lexical matching with a semantic measure, we form a composite text by concatenating the display name and profile description, then encode it with the E5-multilingual model to obtain 768-dimensional embeddings. We then use an Inverted File Index implemented in Facebook AI Similarity Search (FAISS) to efficiently approximate search for nearest-neighbours ~\autocite{douze2024faiss}. This semantic signal complements the lexical indexes, and we applied all three indexes together in the matching pipeline.

We detail the evaluation of retrieval performance in Table~S2. The LSH-based approach on display names provides the strongest individual signal, while the combined approach yields the best overall performance, retrieving over 87\% of true matches within the top 10 candidates. We retrieved 10 candidates per method, leading to at most 30 potential matches per candidate. We find no improvement in classification beyond retrieving 10 candidates per method. 

This indexing strategy improved query throughput. On average, one semantic query took 139.9 ms ($\pm$52.8) with $99^{th}$ percentile duration being 299.6 ms. The locality sensitive hashing for display names took 23.2 ms ($\pm$26.9) and for the handle 42 ms ($\pm$49.6). 

With our efficient indexing strategy in place, we developed a classification model to determine whether candidate pairs represent the same user across platforms. For the final classification of potential pairs, we fine-tuned a DistilBERT model on our labeled dataset. We split the dataset into training (60\%), validation (20\%), and test (20\%) sets, stratified by user to prevent data leakage between splits. This ensures that different profile pairs from the same user do not appear in both the training and validation/test sets, providing a more realistic assessment of generalization. The model demonstrates robust performance across both matching and non-matching classes, with a macro-F1 score of 0.96.

\subsection{Political expression analysis}
For each of the political, economic, and social topics, we manually build keyword dictionaries that capture the diverse terminology academics use on Twitter. We selected the 14 focal topics ex ante, before inspecting any migration outcomes. We started from survey evidence on issue salience and partisan divisions in contemporary U.S. politics from \href{https://news.gallup.com/poll/1675/most-important-problem.aspx}{Gallup} and the \href{https://www.pewresearch.org/politics/2019/12/17/in-a-politically-polarized-era-sharp-divides-in-both-partisan-coalitions/}{Pew Research Center}, which consistently highlight topics such as health care, abortion, climate policy, racial inequality, economic inequality, and free speech as among the most important and polarizing issues. To this set we added platform- and technology-related topics (Twitter/X, artificial intelligence, big tech) that have been especially salient over our study period and are central to online political discussion.

For each topic we constructed a compact keyword dictionary by (i) seeding it with terms used in survey questions and media coverage, (ii) extending this list with common variants, abbreviations, and hashtags observed in a random subsample of our Twitter corpus, and (iii) manually pruning ambiguous terms that generated many false positives in pilot searches. We flagged a user as having "political expression" on a given topic if at least one of their posts contained a dictionary term. We then aggregate these binary indicators to the author level and examine how simply mentioning each topic—regardless of sentiment—predicts the probability of migrating to Bluesky. The keyword dictionary is in Table~S4.

\subsection{Natural experiment}
We now focus on a daily panel of 239{,}917 non-Brazilian Twitter users covering the 14 days before and the 30 days after the Brazilian Supreme Court’s August 30, 2024 decision banning X. We further limit the sample to users that are in the risk set of transitioning, i.e. that had not transitioned yet. To investigate how the shock propagates to not directly affected users, we leverage the fact that some of the non-Brazilian Twitter users have pre-existing network ties to Brazilian users based on our network data compiled before the shock. We identify whether a user is Brazilian/directly affected by the ban based on the academic's affiliation in 2024, as per OpenAlex. For each user, we calculate a time-invariant grouping based on the varying number of Brazilian connections (none, 1–3, and $\geq$ 4), separately for ties to information sources (followees) and audiences (followers). Then, for each day and ego, we define a binary outcome indicating whether ego \(i\) transitions on day $t\in (-14,30)$. To increase the precision of our estimates, we control for the log number of non-Brazilian users that transitioned in the previous period.  We estimate the following event–study specification for followees:

\begin{equation}
y_{it} = 
\sum_{\tau \neq 0} \sum_{g \in \mathcal{G}} 
\beta_{\tau g} \,
\mathbf{1}\{t  = \tau\} \times \mathbf{1}\{G_i = g\}
+ \gamma \log(F_{it})
+ \mu_i + \lambda_t + \lambda_t^{G} + \varepsilon_{it},
\end{equation}

where \(y_{it}\) is an indicator for whether user \(i\) transitions on day \(t\); 
$\tau$ denotes the number of days relative to the ban date; 
\(G_i \in \{0, 1\text{--}3, \geq 4\}\) indexes groups defined by the number of Brazilian followees (with \(G_i = 0\) as the reference category);
\(F_{it}\) is the number of non-Brazilian followers who have transitioned (in logs);
\(\mu_i\) are user fixed effects; 
\(\lambda_t\) are relative-day fixed effects; 
\(\lambda_t^{G}\) are relative-day-by-group follower fixed effects; 
and \(\varepsilon_{it}\) is an idiosyncratic error term. We cluster the standard errors at the user level.  The estimation strategy exploits variation in exposure across groups of users who differ in their number of Brazilian connections. For each treatment dimension (e.g., Brazilian followees), we control for relative-day effects associated with the opposite direction of ties (e.g., Brazilian followers), ensuring that estimated dynamics are not confounded by concurrent exposure through the other channel. Identification relies on a standard two-way fixed effects framework, which recovers the average causal effect under the assumption of parallel trends. Because treatment status is absorbing—once a user becomes exposed, they remain exposed - and non-staggered, the specification does not suffer from the issues commonly discussed in the recent difference-in-differences literature \autocite{dechaisemartinTwowayFixedEffects2023}.

\subsection{Dynamic Propensity Matching Estimates}
For the dynamic matching, we proceed much the same way as for the quasi-experimental event study above, but assemble covariates measured strictly prior to the treatment window (at D-7, one week before the outcome day) and perform same-day matching to estimate risk ratios (RRs). Covariates fall into three categories (Table~S7). Six dynamic saturation measures capture counts and fractions of transitioned followees, followers, and mutuals as of D-7. Ego attributes (75 variables) include country, academic field, gender, political expression, Twitter activity metrics such as follower and following counts and their growth rates, academic productivity measures, and network size. Network composition variables (144 variables) are computed as averages of the same attributes across each ego's audience and information sources. We estimate each design separately for followees, followers, and mutual ties. For approximate nearest-neighbor matching within propensity-score calipers, we use an 11-dimensional subspace consisting of the 6 saturation measures plus 5 network size variables. Diagnostics for covariate balance are reported in Tables~S8 and S9.

Timing treatments indicate whether any neighbor transitioned within the last \(d\) days for \(d \in \{1,\ldots,6\}\) (same-day exposure is excluded). Dose treatments count the number of neighbors who transitioned in the last seven days and bin the count as \(\{0,1,2,3,3+\}\). Placebos include (i) ``future'' timing exposures (events on \(D{+}1\) through \(D{+}6\)) and (ii) permutation placebos that randomly reassign the within-day dose across egos while preserving the dose distribution.

For timing designs, we fit a logistic regression for the treatment indicator; for dose designs, we fit a multinomial logistic model to obtain a generalized propensity score over treatment categories \autocite{rosenbaum1983central,Imbens2000,HiranoImbens2004}. We enforce a caliper of width \(0.1 \times \mathrm{SD}_{\text{pooled}}\) on the propensity-score logit for the treatment. Within each day, we generate candidate controls for each treated ego using approximate nearest neighbors in the standardized covariate space spanned by the pre-treatment saturation measures. We implemented this shortlisting with FAISS using a hierarchical navigable small-world index \autocite{Johnson2017,MalkovYashunin2020}. Among candidates that satisfy the propensity caliper, we select a single control without replacement, preferring smaller standardized Mahalanobis distance computed on the core pre-treatment saturation features when multiple feasible controls remain \autocite{Mahalanobis1936,Rubin1980}. We skipped day–specifications with insufficient treated or control counts. For each day and specification, we compute treated and control risks from the matched pairs and form a daily risk ratio. To stabilize per-day estimates when a cell count is zero, we apply the Haldane–Anscombe continuity correction by adding 0.5 to each cell of the \(2\times2\) table \autocite{Haldane1956,Anscombe1956}. For pooled inference, we sum the \(2\times2\) cell counts across days within a specification and report a Wald confidence interval for \(\log(\mathrm{RR})\) using the Katz method \autocite{Katz1978}. We processed placebos identically.

\subsection{Contagion modelling} \label{sec:method:contagion}
To identify the specific contagion mechanism driving each individual's platform transition, we extend the node-level classification framework of \citeauthor{andres2025} to incorporate exogenous shocks\autocite{andres2025}. Table~\ref{tab:contagion_rules} summarizes the four mechanisms governing transitions. Simple contagion operates through independent influence from each adopted neighbor following probability $P\{A_i(t)=1\} = 1-(1-\beta_i)^{m_i(t-1)}$ where $A_i(t) \in \{0,1\}$ indicates adoption on day $t$ and $m_i(t-1)$ counts followees who transitioned before day $t$ and $\beta_i$ is the transmission rate. Complex contagion requires a threshold fraction of one's network to adopt before transition occurs, formalized as $A_i(t)=\mathbbm{1}[m_i(t)/k_i \ge \phi_i]$ for $k_i > 0$, where $k_i$ is in-degree and $\phi_i$ is the individual threshold; for nodes with $k_i = 0$, complex contagion cannot occur since $m_i(t) = 0$ by definition. Spontaneous adoption occurs at constant rate $r$ independent of network exposure. Shock-driven adoption responds to external events through time-varying intensity.

\begin{equation}
  \lambda_{\text{shock}}(t) = \sum_{j=1}^{5} \gamma_j (t - \tau_j + 1)^{-\alpha_j}
  \cdot \mathbf{1}\!\bigl[\tau_j \le t < \tau_{j+1}\bigr],
  \label{eq:shock}
\end{equation}

where $\tau_j$ marks the peak day of shock $j$, $\gamma_j$ denotes its relative height normalized to the largest shock, and $\alpha_j$ represents the shock-specific power-law decay exponent\autocite{sornette2004,Crane2008}. We fit individual decay exponents to each burst using log-log regression (see Table~S11). As an independent validation, we apply the degree-adoption order correlation test\autocite{cencettiDistinguishingSimpleComplex2023}. Under simple contagion high-degree nodes are exposed earlier and adopt first ($\rho < 0$), whereas complex contagion weakens or reverses this pattern. 

\begin{table}[!bh]
\centering
\begin{tabularx}{\linewidth}{@{}lXc@{}}
\toprule
Mechanism & Adoption rule (daily probability / trigger) & Parameter Range\\
\midrule
Simple (Sm) &
$\displaystyle
  P\bigl\{A_i(t)=1\bigr\}
  =1-\bigl(1-\beta_i\bigr)^{m_i(t)}
$ &
$\beta_i\in(0.011, 1.0)$ \\[6pt]
Complex (Cx) &
$\displaystyle
  A_i(t)=\mathbbm 1\!\Bigl[
    \tfrac{m_i(t)}{k_i}\ge\phi_i
  \Bigr]
$ &
$\phi_i\in(0.001, 0.246)$ \\[6pt]
Spontaneous (St) &
$\displaystyle
  P\bigl\{A_i(t)=1\bigr\}=r
$ &
$r=60\times10^{-6}$ \\[6pt]
Shock (Sk) &
$\displaystyle
  P\bigl\{A_i(t)=1\bigr\}=\lambda_{\text{shock}}(t)
$ &
$\{\gamma_j,\alpha_j\}_{j=1}^{5},\;\bar{\alpha}=0.59$ \\\bottomrule
\end{tabularx}
\caption{Local and global models of Twitter-to-Bluesky transitions. $m_i(t)$ = number of transitioned \emph{followees} of user~\(i\) on day~\(t\);
\(k_i\) = in-degree.}
\label{tab:contagion_rules}
\end{table}

We derived all simulation parameters from empirical adoption patterns of users who transitioned outside shock periods to isolate organic diffusion dynamics. We derived simple contagion transmission rates by taking the reciprocal of observed exposure counts, yielding a distribution with mean $\beta=0.089$ and range $(0.011, 1.0)$. We set complex contagion thresholds to the fraction of exposed neighbors for the same population subset, producing values with mean
$\phi=0.146$ and range $(0.001, 0.246)$. We calculated the spontaneous adoption rate as the ratio of zero-exposure adopters to total susceptible-days across all nodes at $r=60\times10^{-6}$ per day. We derived activity levels from Twitter posting patterns spanning 2022-2025,
log-transformed and normalized to obtain realistic heterogeneity with mean activity of 0.032.

We executed 100 independent realizations of the mixed-contagion model on the entire directed Twitter follower network. Each run terminates when 18\% of nodes adopt or two simulation 
years elapse. This procedure generates 5.2 million unique node-transition events after removing duplicates to ensure class balance. For each transition event (user $u$ adopting on day $t_u$), we computed seven egocentric features measured at adoption time. Active influences count the number of infected followees $m = |\mathcal{N}^-(u) \cap \mathcal{I}(t_u)|$. Network degree equals in-degree $k = |\mathcal{N}^-(u)|$. Peer saturation measures the share of infected followees as $m/k$. Exposure duration calculates time since first exposure $t_u - \min_{v\in\mathcal{N}^-(u)\cap\mathcal{I}(t_u)}t_v$. Influence recency computes time since last exposure. Shock intensity records the instantaneous value $\lambda_{\text{shock}}(t_u)$. Shock recency measures days since the most recent shock peak.

We trained both XGBoost and Random Forest classifiers on these synthetic events. XGBoost achieves superior performance with macro-averaged F1 score of 0.70 compared with 0.69 for Random Forest. Feature importance analysis reveals distinct patterns across mechanisms (Figure~S1). We assess importance using GAIN, the average reduction in loss from splits using each feature \autocite{chen2016xgboost}, and direction of influence using Local Interpretable Model-agnostic Explanations (LIME), which fits local linear approximations around each prediction \autocite{ribeiro2016should}. Simple contagion depends primarily on active influences and exposure duration. Complex contagion shows clear dependence on peer saturation. Spontaneous adoption exhibits relatively uniform feature importance. Shock-driven transitions concentrate weight on shock intensity and recency.

\section{Data and code availability}
In accordance with our ethical approval, we do not publicly share raw data that allows users to be identified. These materials will be deposited on Zenodo upon acceptance (the DOI and URL will be added here) and may be made available from the authors upon reasonable request via Zenodo. 
All codes will be made publicly available at the same location.


\section{Author contributions}
F.D., D.Q., P.G., and A.B. conceived the project and developed the theory and methodology. D.Q., P.G., and F.D. procured and cleaned the data. D.Q. and F.D. performed the computations. A.B. supervised the work. D.Q. and F.D. wrote the manuscript with support from A.B. and P.G. All authors discussed the results and contributed to the final manuscript.

\section{Ethics declarations}
The authors declare no competing interests. 
The study was approved by Departmental Review as part of the ethics review process of the London School of Economics and Political Science (Request Number 474184).

\newpage
\setcounter{section}{0}
\renewcommand{\thesection}{S\arabic{section}}
\setcounter{table}{0}
\setcounter{figure}{0}
\captionsetup[figure]{labelformat=sfiglab}
\captionsetup[table]{labelformat=stablab}

\begin{center}
{\Large\bfseries Supplementary Information}\\[0.5em]
{\large Simple contagion drives population-scale platform migration}
\end{center}
\vspace{1em}

\section{Cross-platform identity matching}

\begin{table}[H]
\centering
\begin{tabular}{lrrrr}
\toprule
\textbf{Class} & \textbf{Precision} & \textbf{Recall} & \textbf{F1-score} & \textbf{Support} \\
\midrule
non-matching & 0.95 & 0.98 & 0.97 & 106 \\
matching & 0.97 & 0.93 & 0.95 & 57 \\
\midrule
Macro avg    & 0.96 & 0.96 & 0.96 & 163 \\
\midrule
Accuracy     & \multicolumn{3}{r}{0.96} & 163 \\
\bottomrule
\end{tabular}
\caption{Test set performance for the entity matching model.}
\label{tab:S_test_results_config1}
\end{table}

\begin{table}[H]
    \centering
    \begin{tabular}{ccccc}
    \toprule
    \(k\) & Semantic & LSH-Display & LSH-Handle & Combined \\
    \midrule
    1  & 4.5\% & 7.0\% & 6.3\% & 13.1\% \\
    3  & 13.9\% & 25.4\% & 17.4\% & 34.0\% \\
    5  & 20.2\% & 41.8\% & 27.9\% & 54.3\% \\
    10 & 37.6\% & 77.0\% & 52.3\% & 87.2\% \\
    30 & 37.6\% & 85.7\% & 55.1\% & 91.8\% \\
    100 & 37.6\% & 87.1\% & 55.1\% & 93.3\% \\
    \bottomrule
    \end{tabular}
    \caption{Cross-platform account identification retrieval performance per index. Each value represents the percentage of true positives successfully retrieved within the top-\(k\) candidates using different matching methods (287 ground truth pairs). At $k{=}10$, unique contributions are: LSH-Display 24.0\%, LSH-Handle 5.2\%, Semantic 1.4\%.}
    \label{tab:S_blockingtable}
\end{table}

\section{Sample characteristics} \label{sec:S_sample_descriptives_external_validity}

\begin{table}[H]
\centering
\caption{Twitter deletion and inactivity by transition status.}
\begin{tabular}{p{0.3\linewidth}p{0.2\linewidth}p{0.2\linewidth}p{0.2\linewidth}}
\toprule
 & \textbf{Not Transitioned} & \textbf{Transitioned} & \textbf{Total} \\
\midrule
\multicolumn{4}{l}{\textit{Deletion by 2025}} \\
Deleted   & 11.5\% (26,026)  & 18.6\% (9,239)  & 12.8\% (35,265) \\
Not deleted & 88.5\% (200,564) & 81.4\% (40,492) & 87.2\% (241,056) \\
\midrule
\multicolumn{4}{l}{\textit{Activity until 2025}} \\
Inactive & 31.4\% (62,922) & 17.9\% (7,235) & 29.1\% (70,157) \\
Active   & 68.6\% (137,642) & 82.1\% (33,257) & 70.9\% (170,899) \\
\bottomrule
\end{tabular}
\caption*{\small Cross-tabulation of Twitter deletion and activity status by Bluesky transition. Transitioned users were more likely to have deleted their Twitter account by early 2025.}
\label{tab:S_transition_deletion_activity}
\end{table}

\section{Political expression analysis}

\begin{table}[H]
\centering
\footnotesize
\begin{tabular}{@{}p{3.8cm}p{10.8cm}@{}}
\toprule
\textbf{Topic} & \textbf{Keywords (examples)} \\
\midrule
Healthcare \& Public Health
& public health, universal health care, healthcare reform, NHS, single-payer, healthcare funding, Obamacare, Affordable Care Act \\
\addlinespace
Abortion Rights
& abortion, reproductive rights, pro-choice, pro-life, fetus, Planned Parenthood, heartbeat bill, Roe v. Wade \\
\addlinespace
Climate \& Renewables
& climate, global warming, renewables, solar, wind, green energy, fossil fuels, net zero, emissions, carbon tax, Paris Agreement \\
\addlinespace
Donald Trump
& Trump, MAGA, Big Lie, Donald \\
\addlinespace
Elon Musk
& Elon, Musk, Tesla, SpaceX, Neuralink, DOGE \\
\addlinespace
Free Speech \& Censorship
& free speech, censorship, deplatform, content moderation, misinformation, fake news \\
\addlinespace
Artificial Intelligence
& artificial intelligence, AI, machine learning, ChatGPT, OpenAI, Anthropic \\
\addlinespace
Joe Biden
& Biden, Bidenomics, "Sleepy Joe" \\
\addlinespace
Twitter/X
& Twitter, tweet, X \\
\addlinespace
Economic Liberalism
& free market, capitalism, deregulation, tax cuts, privatization, free trade, globalization \\
\addlinespace
Economic Growth
& economic growth, GDP growth, productivity, innovation, degrowth \\
\addlinespace
Economic Inequality
& income inequality, wealth gap, redistribution, minimum wage, social safety net \\
\addlinespace
Big Tech \& Gov't Influence
& big tech, regulatory capture, Mark Zuckerberg, venture capital \\
\addlinespace
Affirmative Action
& affirmative action, DEI, meritocracy, systemic racism, reverse discrimination \\
\bottomrule
\end{tabular}
\caption{Political topics and associated keyword dictionaries used to code users' political expression.}
\label{tab:S_political_keywords}
\end{table}

\begin{table}[]
\begin{tabular}{lllll}
\toprule
Theme & Topic & $\hat{\beta}$ & SE & $p$ \\
\midrule
\textbf{Economy} & Economic Growth & -0.009 & 0.002 & 0.000 \\
 & Economic Inequality & 0.029 & 0.003 & 0.000 \\
 & Economic Liberalism & 0.007 & 0.003 & 0.012 \\
\textbf{Environment} & Climate \& Renewables & 0.015 & 0.002 & 0.000 \\
\textbf{Politics \& Social Issues} & Abortion Rights & 0.035 & 0.003 & 0.000 \\
 & Affirmative Action & 0.047 & 0.003 & 0.000 \\
 & Free Speech & 0.028 & 0.003 & 0.000 \\
 & Healthcare & -0.007 & 0.002 & 0.002 \\
\textbf{Social Media} & Twitter & 0.052 & 0.002 & 0.000 \\
\textbf{Technology \& Business} & Artificial Intelligence & 0.047 & 0.002 & 0.000 \\
 & Big Tech & 0.016 & 0.003 & 0.000 \\
 & Elon Musk & 0.012 & 0.003 & 0.000 \\
\textbf{US Politics} & Donald Trump & -0.000 & 0.002 & 0.981 \\
 & Joe Biden & 0.025 & 0.003 & 0.000 \\
\textbf{Control} & Board And Card Games & 0.007 & 0.002 & 0.001 \\
 & Coffee And Tea Preferences & 0.013 & 0.002 & 0.000 \\
 & Cooking And Recipes & 0.017 & 0.002 & 0.000 \\
 & Gardening And Houseplants & 0.011 & 0.003 & 0.000 \\
 & Hiking And Outdoor Activities & 0.011 & 0.003 & 0.000 \\
 & Hobbies And Crafts & 0.022 & 0.002 & 0.000 \\
 & Music Tastes And Genres & 0.016 & 0.002 & 0.000 \\
 & Pets And Animal Companions & 0.015 & 0.002 & 0.000 \\
 & Photography And Photo Sharing & -0.008 & 0.002 & 0.000 \\
 & Sports And Fitness & -0.008 & 0.002 & 0.000 \\
 & Travel And Tourism & -0.000 & 0.002 & 0.883 \\
\bottomrule
\end{tabular}
\caption{Linear Probability Model coefficients for effect of expression on political \& social on transition probability from Twitter to Bluesky.}
\label{tab:S_lpm_full_table_expression}
\end{table}

\begin{table}[htbp]
\centering
\begin{adjustbox}{width=\textwidth,center}
\begin{tabular}{lr lr lr}
\toprule
 & & & & & \\
\midrule
Intercept & -0.075***\phantom{} & GB (country) & \phantom{-}0.047***\phantom{} & Economics (field) & \phantom{-}0.023***\phantom{} \\
Active Account & \phantom{-}0.094***\phantom{} & IE (country) & -0.004\phantom{***} & Energy (field) & -0.013\phantom{***} \\
Gender Unclear & -0.003\phantom{***} & IN (country) & -0.039***\phantom{} & Engineering (field) & -0.026***\phantom{} \\
Male & \phantom{-}0.025***\phantom{} & IT (country) & -0.005\phantom{***} & Environmental Sci. (field) & \phantom{-}0.004\phantom{***} \\
Post Bluesky Tweets (000s) & \phantom{-}0.000\phantom{***} & NL (country) & \phantom{-}0.064***\phantom{} & Health Professions (field) & -0.036***\phantom{} \\
Pre Bluesky Tweets (000s) & \phantom{-}0.002***\phantom{} & Other (country) & \phantom{-}0.001\phantom{***} & Immunology (field) & \phantom{-}0.016*\phantom{**} \\
Race Unclear & \phantom{-}0.029***\phantom{} & SE (country) & \phantom{-}0.068***\phantom{} & Materials Science (field) & -0.026***\phantom{} \\
White & \phantom{-}0.050***\phantom{} & US (country) & \phantom{-}0.036***\phantom{} & Mathematics (field) & \phantom{-}0.033**\phantom{*} \\
BE (country) & \phantom{-}0.029***\phantom{} & Arts \& Humanities (field) & \phantom{-}0.066***\phantom{} & Medicine (field) & -0.037***\phantom{} \\
BR (country) & \phantom{-}0.077***\phantom{} & Biochemistry (field) & \phantom{-}0.022***\phantom{} & Neuroscience (field) & \phantom{-}0.042***\phantom{} \\
CA (country) & \phantom{-}0.017***\phantom{} & Business (field) & -0.014*\phantom{**} & Nursing (field) & -0.014\phantom{***} \\
CH (country) & \phantom{-}0.069***\phantom{} & Chemical Eng. (field) & -0.020\phantom{***} & Pharmacology (field) & -0.038*\phantom{**} \\
DE (country) & \phantom{-}0.102***\phantom{} & Chemistry (field) & -0.015\phantom{***} & Physics (field) & \phantom{-}0.004\phantom{***} \\
DK (country) & \phantom{-}0.075***\phantom{} & Computer Science (field) & -0.003\phantom{***} & Psychology (field) & \phantom{-}0.005\phantom{***} \\
ES (country) & \phantom{-}0.024***\phantom{} & Decision Sciences (field) & \phantom{-}0.001\phantom{***} & Social Sciences (field) & \phantom{-}0.038***\phantom{} \\
FI (country) & \phantom{-}0.065***\phantom{} & Dentistry (field) & -0.023\phantom{***} & Veterinary (field) & \phantom{-}0.017\phantom{***} \\
FR (country) & \phantom{-}0.057***\phantom{} & Earth Sciences (field) & \phantom{-}0.030***\phantom{} &  &  \\
\midrule
$R^2$ = 0.120 & & Adj. $R^2$ = 0.120 & & $N$ = 206,898 & \\
\bottomrule
\multicolumn{6}{l}{\textit{Note:} * $p<0.05$, ** $p<0.01$, *** $p<0.001$.} \\
\end{tabular}
\end{adjustbox}
\caption{Linear Probability Model coefficients for demographic, country, and field controls on transition probability from Twitter to Bluesky.}
\label{tab:S_lpm_full_table_controls}
\end{table}

\section{Dynamic matching diagnostics}
\label{sec:S_app_dynamic}

\begin{table}[htbp]
\centering
\caption{Covariates Used in Propensity Score Estimation}
\label{tab:S_psm_variables}
\begin{threeparttable}
\begin{tabular}{lp{9cm}r}
\toprule
Category & Variables & Count \\
\midrule
\textbf{Dynamic Saturation}$^\dagger$ & Count and fraction of transitioned followees, followers, and mutuals (measured at D-7) & 6 \\
\midrule
\textbf{Ego Attributes} & Country (19); Academic field (20); Gender (4); Political content (18); Activity metrics: follower count, following count, statuses count, favourites count (each with 2025 level and growth); Academic metrics: cited\_by\_count, impact\_factor, works\_count; Network size: n\_audience, n\_info\_sources, total\_degree$^\dagger$ & 75 \\
\midrule
\textbf{Network Composition}$^*$ & Same attribute set as Ego Attributes above, excluding network size, computed separately for audience (followers) and information sources (followees) & 144 \\
\midrule
\multicolumn{2}{l}{\textbf{Total Covariates}} & \textbf{225} \\
\bottomrule
\end{tabular}
\begin{tablenotes}
\small
\item[$\dagger$] Used in ANN matching (11 dimensions: 6 saturation + 5 network size variables).
\item[$*$] Computed as network-averaged values across each ego's followers and followees.
\end{tablenotes}
\end{threeparttable}
\end{table}

This section reports diagnostics for the dynamic propensity score matching procedure described in Methods. We report four metrics.

The generalized propensity score (GPS) area under the ROC curve (AUC) measures how well the propensity model distinguishes between treatment levels. Higher AUC values indicate that treated and control groups have meaningfully different covariate profiles before matching, validating the need for adjustment. Values near 0.5 would suggest treatment is essentially random conditional on covariates.

The standardized difference in the propensity score logit ($\Delta$ logit) after applying the caliper measures residual imbalance between matched treated and control units. Values close to zero indicate that the caliper successfully eliminates large propensity score differences, ensuring treated and control units are comparable in their predicted probability of treatment.

The approximate nearest-neighbor (ANN) distance measures the Euclidean distance in standardized covariate space between each treated unit and its matched control. Smaller distances indicate that matched pairs are similar not just in their propensity scores but also in the underlying covariates themselves, reducing the risk of bias from residual imbalance.

Overlap is defined as the fraction of treated units for which at least one control unit falls within the propensity score caliper. Values close to one indicate good common support, meaning that for nearly all treated units we can find comparable controls. Low overlap would indicate extrapolation beyond the support of the data.

Table~\ref{tab:S_diag_dose_compact2} reports diagnostics for the dose design, in which treatment is defined as the number of neighbors who transitioned to Bluesky in the past week, binned as \{1, 2, 3, 3+\}. The GPS AUC increases monotonically with dose level, ranging from approximately 0.70 at dose 1 to above 0.97 at dose 3+. This pattern indicates that users with higher exposure levels are increasingly distinguishable from unexposed controls, which is expected since heavy exposure is relatively rare. Post-caliper balance is uniformly tight across all dose levels, with median $\Delta$ logit values below 0.055 in all cases, confirming that the 0.10 caliper effectively eliminates propensity score imbalance. ANN distances increase modestly with dose level but remain moderate throughout, indicating that matched pairs remain reasonably similar even at high exposure levels. Overlap is essentially perfect (median = 1.000) at doses 1 through 3, with some erosion at the highest dose category where treated units are rarer and more extreme. Even at dose 3+, overlap remains above 0.76 for all tie directions, indicating adequate common support.

Table~\ref{tab:S_diag_time_compact2} reports diagnostics for the timing design, in which treatment is defined as whether any neighbor transitioned within the last $d$ days, for $d$ ranging from 1 to 6. The GPS AUC is uniformly high across all lags, ranging from 0.88 to 0.92 depending on tie direction. This consistency reflects the fact that the timing treatment is binary (any recent exposure vs. none), producing stable discrimination across different lag windows. Post-caliper balance is controlled throughout, with median $\Delta$ logit values between 0.10 and 0.19. These values are somewhat larger than in the dose design because the binary treatment definition creates coarser groupings, but they remain well within acceptable bounds. ANN distances are stable across lags at approximately 0.26 to 0.36, and overlap is near-perfect in all specifications (median $\geq$ 0.999).

Together, these diagnostics demonstrate that our matching procedure successfully balances treated and control groups across both dose and timing specifications, supporting valid causal inference from the matched comparisons reported in the main text.

\begin{table}[ht]
\centering
\footnotesize
\caption{Dose diagnostics (Level = bin; caliper = 0.10)}
\label{tab:S_diag_dose_compact2}
\begin{tabular}{lrrrrrrrrrr}
\toprule
 &  &  &  & \multicolumn{2}{c}{auc} & \multicolumn{2}{c}{$\Delta$logit} & \multicolumn{2}{c}{ANN dist} & \multicolumn{1}{c}{overlap} \\
\cmidrule(lr){5-6} \cmidrule(lr){7-8} \cmidrule(lr){9-10} \cmidrule(l){11-11}
Level & days & treated & controls & \scriptsize median & \scriptsize min & \scriptsize p50 & \scriptsize p90 & \scriptsize p50 & \scriptsize p90 & \scriptsize median \\
\midrule
\multicolumn{11}{l}{\textbf{Followee}}\\
 1 & 670 & 14,110,855 & 157,950,246 & 0.703 & 0.403 & 0.025 & 0.064 & 0.125 & 0.439 & 1.000 \\
 2 & 615 & 4,621,841 & 157,950,246 & 0.849 & 0.490 & 0.028 & 0.068 & 0.183 & 0.674 & 1.000 \\
 3 & 477 & 2,204,538 & 157,950,246 & 0.930 & 0.521 & 0.032 & 0.069 & 0.227 & 0.883 & 1.000 \\
 3+ & 416 & 4,405,531 & 157,950,246 & 0.970 & 0.763 & 0.054 & 0.099 & 0.305 & 1.054 & 0.987 \\
\midrule
\multicolumn{11}{l}{\textbf{Follower}}\\
 1 & 670 & 13,138,809 & 160,997,845 & 0.684 & 0.477 & 0.015 & 0.039 & 0.113 & 0.516 & 1.000 \\
 2 & 615 & 3,924,538 & 160,997,845 & 0.859 & 0.505 & 0.020 & 0.048 & 0.209 & 0.945 & 0.999 \\
 3 & 506 & 1,787,476 & 160,997,845 & 0.948 & 0.725 & 0.027 & 0.059 & 0.315 & 1.244 & 0.996 \\
 3+ & 452 & 3,444,343 & 160,997,845 & 0.987 & 0.758 & 0.054 & 0.106 & 0.408 & 1.567 & 0.761 \\
\midrule
\multicolumn{11}{l}{\textbf{Mutual}}\\
 1 & 669 & 8,435,726 & 170,233,975 & 0.718 & 0.500 & 0.025 & 0.063 & 0.137 & 0.509 & 1.000 \\
 2 & 530 & 2,260,431 & 170,233,975 & 0.877 & 0.509 & 0.025 & 0.064 & 0.218 & 0.866 & 1.000 \\
 3 & 401 & 944,416 & 170,233,975 & 0.953 & 0.561 & 0.030 & 0.064 & 0.307 & 1.192 & 0.997 \\
 3+ & 331 & 1,418,463 & 170,233,975 & 0.981 & 0.776 & 0.052 & 0.091 & 0.439 & 1.530 & 0.896 \\
\bottomrule
\end{tabular}
\end{table}

\begin{table}[t]
\centering
\footnotesize
\caption{Timing diagnostics (Level = lag; caliper = 0.10)}
\label{tab:S_diag_time_compact2}
\begin{tabular}{lrrrrrrrrrr}
\toprule
 &  &  &  & \multicolumn{2}{c}{auc} & \multicolumn{2}{c}{$\Delta$logit} & \multicolumn{2}{c}{ANN dist} & \multicolumn{1}{c}{overlap} \\
\cmidrule(lr){5-6} \cmidrule(lr){7-8} \cmidrule(lr){9-10} \cmidrule(l){11-11}
Level & days & treated & controls & \scriptsize median & \scriptsize min & \scriptsize p50 & \scriptsize p90 & \scriptsize p50 & \scriptsize p90 & \scriptsize median \\
\midrule
\multicolumn{11}{l}{\textbf{Followee}}\\
 1 & 630 & 7,523,114 & 163,044,976 & 0.906 & 0.848 & 0.148 & 0.264 & 0.339 & 1.004 & 1.000 \\
 2 & 647 & 12,525,388 & 162,645,333 & 0.899 & 0.850 & 0.160 & 0.285 & 0.308 & 0.929 & 1.000 \\
 3 & 656 & 16,483,858 & 161,123,550 & 0.899 & 0.852 & 0.169 & 0.297 & 0.292 & 0.892 & 1.000 \\
 4 & 664 & 19,820,290 & 159,953,062 & 0.899 & 0.855 & 0.175 & 0.307 & 0.275 & 0.859 & 1.000 \\
 5 & 670 & 22,737,296 & 158,660,514 & 0.901 & 0.852 & 0.181 & 0.317 & 0.264 & 0.830 & 1.000 \\
 6 & 675 & 25,343,548 & 157,407,977 & 0.900 & 0.850 & 0.189 & 0.329 & 0.256 & 0.804 & 1.000 \\
\midrule
\multicolumn{11}{l}{\textbf{Follower}}\\
 1 & 617 & 6,369,476 & 160,678,955 & 0.892 & 0.809 & 0.100 & 0.177 & 0.361 & 1.323 & 0.999 \\
 2 & 645 & 10,715,962 & 163,913,273 & 0.881 & 0.814 & 0.098 & 0.174 & 0.331 & 1.183 & 0.999 \\
 3 & 657 & 14,226,061 & 163,652,090 & 0.878 & 0.818 & 0.099 & 0.175 & 0.311 & 1.102 & 0.999 \\
 4 & 665 & 17,233,342 & 162,810,753 & 0.876 & 0.818 & 0.102 & 0.178 & 0.298 & 1.039 & 0.999 \\
 5 & 670 & 19,892,695 & 161,505,115 & 0.875 & 0.819 & 0.102 & 0.178 & 0.283 & 0.992 & 0.999 \\
 6 & 675 & 22,296,331 & 160,455,194 & 0.876 & 0.818 & 0.104 & 0.181 & 0.273 & 0.958 & 0.999 \\
\midrule
\multicolumn{11}{l}{\textbf{Mutual}}\\
 1 & 612 & 3,359,221 & 162,335,495 & 0.917 & 0.847 & 0.139 & 0.249 & 0.362 & 1.113 & 1.000 \\
 2 & 640 & 5,884,132 & 167,391,388 & 0.906 & 0.844 & 0.149 & 0.265 & 0.335 & 1.021 & 1.000 \\
 3 & 652 & 8,001,883 & 168,522,553 & 0.903 & 0.845 & 0.156 & 0.278 & 0.320 & 0.978 & 1.000 \\
 4 & 661 & 9,859,428 & 169,101,695 & 0.901 & 0.847 & 0.164 & 0.291 & 0.308 & 0.937 & 1.000 \\
 5 & 668 & 11,530,274 & 169,326,050 & 0.901 & 0.847 & 0.168 & 0.297 & 0.299 & 0.916 & 1.000 \\
 6 & 674 & 13,060,008 & 169,420,774 & 0.901 & 0.847 & 0.172 & 0.307 & 0.291 & 0.899 & 1.000 \\
\bottomrule
\end{tabular}
\end{table}

\newpage
\section{Shock events}

\begin{table}[htbp]
    \centering
    \begin{tabular}{lrrp{8cm}}
    \hline
    \textbf{Period} & \textbf{Count} & \textbf{Percentage} & \textbf{Context} \\
    \hline
    2023-07-01 - 2023-07-03 & 473 & 0.95\% & Twitter implemented restrictions, including mandatory login and daily viewing limits, making the platform nearly unusable. \autocite{Conger2023} \\
    2023-09-19 - 2023-09-20 & 1,016 & 2.04\% & Newly renamed X, its leadership announced potential monthly subscription fees for all users. \autocite{Milmo2023} \\
    2024-02-06 - 2024-02-07 & 417 & 0.84\% & Bluesky made its official public debut, opening registrations to everyone after invite-only beta testing. \autocite{Silberling2024} \\
    2024-08-31 & 153 & 0.31\% & Brazil suspended access to X following a judicial order amid a dispute with the platform's ownership. 38\% of transitioned users had Brazilian affiliation on this date. \autocite{Rogero2024} \\
    2024-11-06 - 2024-11-14 & 6,077 & 12.21\% & Significant surge in Bluesky sign-ups following the US presidential election. \autocite{Ittimani2024} \\
    \hline
    All shock periods & 8,136 & 16.35\% &  \\
    Non-shock periods & 41,618 & 83.65\% & \\
    \hline
    \end{tabular}
    \caption{Twitter/X to Bluesky Migration Events. Shocks are defined as at least 150 academics transitioning on one day and the daily transition number exceeding the 30 day moving average by more than 3 standard deviations.}
    \label{tab:S_migration-events}
\end{table}

\begin{table}[htbp]
\centering
\begin{tabular}{lcccc}
\toprule
Shock Event & Peak Count & Relative Height & $\hat{\alpha}$ & \textbf{$R^2$} \\
\midrule
Outage on Twitter        & 322  & 0.183 & 0.626 & 0.871 \\
X fees             & 590  & 0.335 & 0.231 & 0.585 \\
Public launch      & 247  & 0.140 & 0.775 & 0.890 \\
Brazil blocks X    & 153  & 0.087 & 0.556 & 0.885 \\
US election        & 1759 & 1.000 & 0.679 & 0.798 \\
\bottomrule
\end{tabular}
\caption{Power-law decay parameters for five major migration shock events. Shocks are defined as days with $>150$ academic transitions exceeding the 30-day moving average by $>3$ standard deviations. Power-law exponents ($\hat{\alpha}$) estimated from post-peak decay patterns following $\lambda_{\text{shock}}(t) = \gamma_j(t-\tau_j)^{-\alpha}$, where $\tau_j$ is the peak day and $\gamma_j$ is the relative shock height (normalized to US election $= 1.0$). Exponents estimated via robust regression (Huber loss) to reduce sensitivity to outlier days in the post-shock decay tail.}
\label{tab:S_shock_summary}
\end{table}

\section{Contagion mechanism classification}

To validate that our classifier distinguishes mechanisms based on theoretically meaningful features, we examine feature importance using two complementary approaches. GAIN measures the average reduction in loss from splits using each feature, capturing overall predictive importance \autocite{chen2016xgboost}. Local Interpretable Model-agnostic Explanations (LIME) fits local linear approximations around each prediction, revealing the direction and magnitude of each feature's influence on classification \autocite{ribeiro2016should}.

\begin{figure}[ht]
  \centering
  \includegraphics[width=\textwidth]{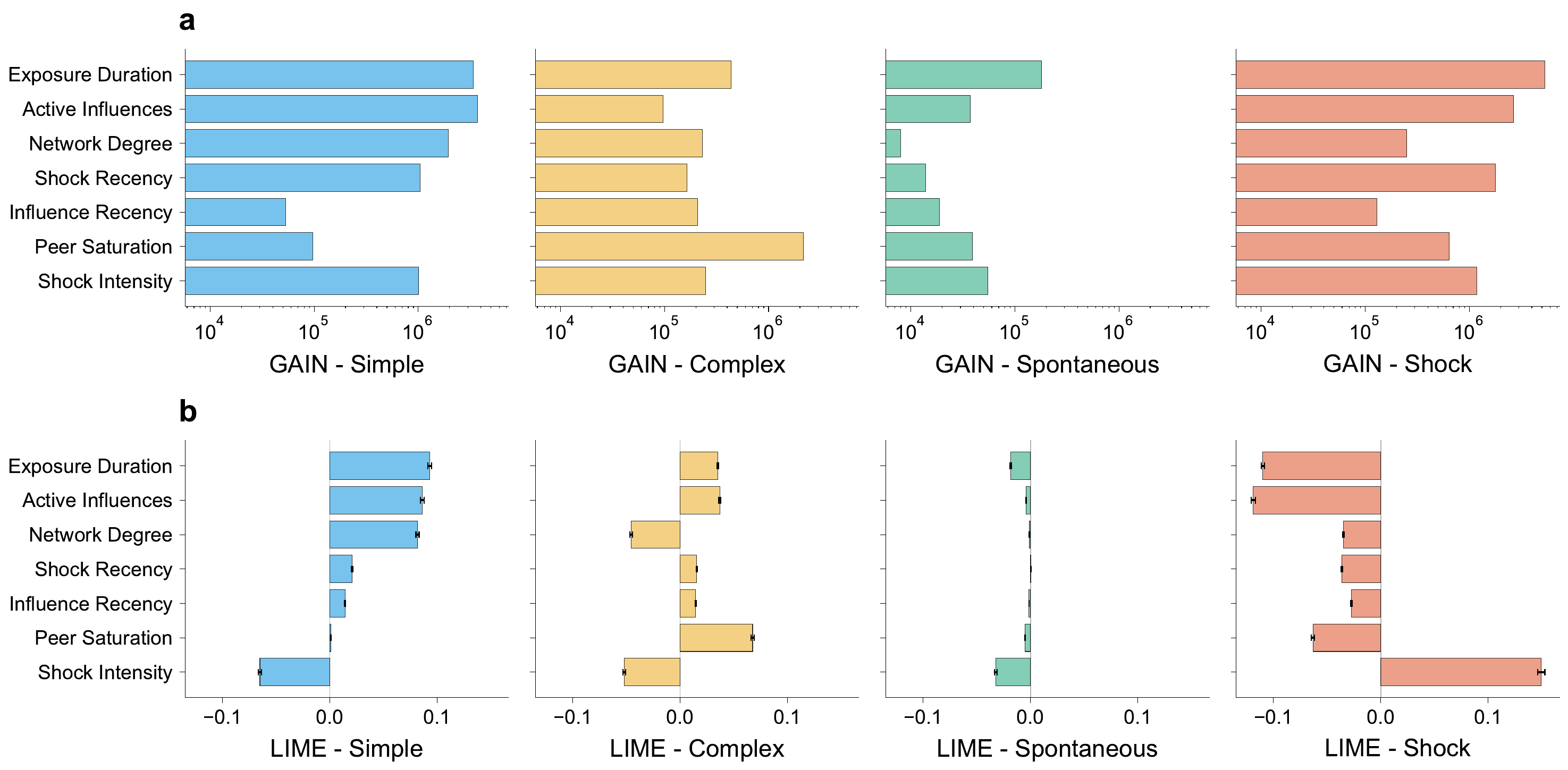}
  \caption{\textbf{Feature importance for contagion mechanism classification.}
  \textbf{a} GAIN scores from XGBoost, measuring the average reduction in loss from splits using each feature.
  \textbf{b} LIME coefficients showing direction and magnitude of local linear effects for each mechanism.
  Simple contagion (blue) depends on exposure duration and active influences. Complex contagion (yellow) relies on peer saturation. Spontaneous adoption (green) shows relatively uniform importance. Shock-driven transitions (red) concentrate on shock intensity and recency.}
  \label{fig:S_gain_lime}
\end{figure}

Figure~\ref{fig:S_gain_lime} confirms that each mechanism exhibits the expected feature signature. For simple contagion, exposure duration and active influences dominate, consistent with independent per-contact transmission where longer exposure and more infected contacts increase adoption probability. Complex contagion shows elevated importance for peer saturation, reflecting the threshold-based trigger where the fraction of adopted neighbors matters more than raw counts. Spontaneous adoption displays relatively uniform feature importance with no single dominant predictor, as expected for a mechanism independent of network exposure. Shock-driven transitions concentrate importance on shock intensity and shock recency, confirming that external events rather than network structure drive these adoptions.

\begin{figure}[ht]
  \centering
  \includegraphics[width=\textwidth]{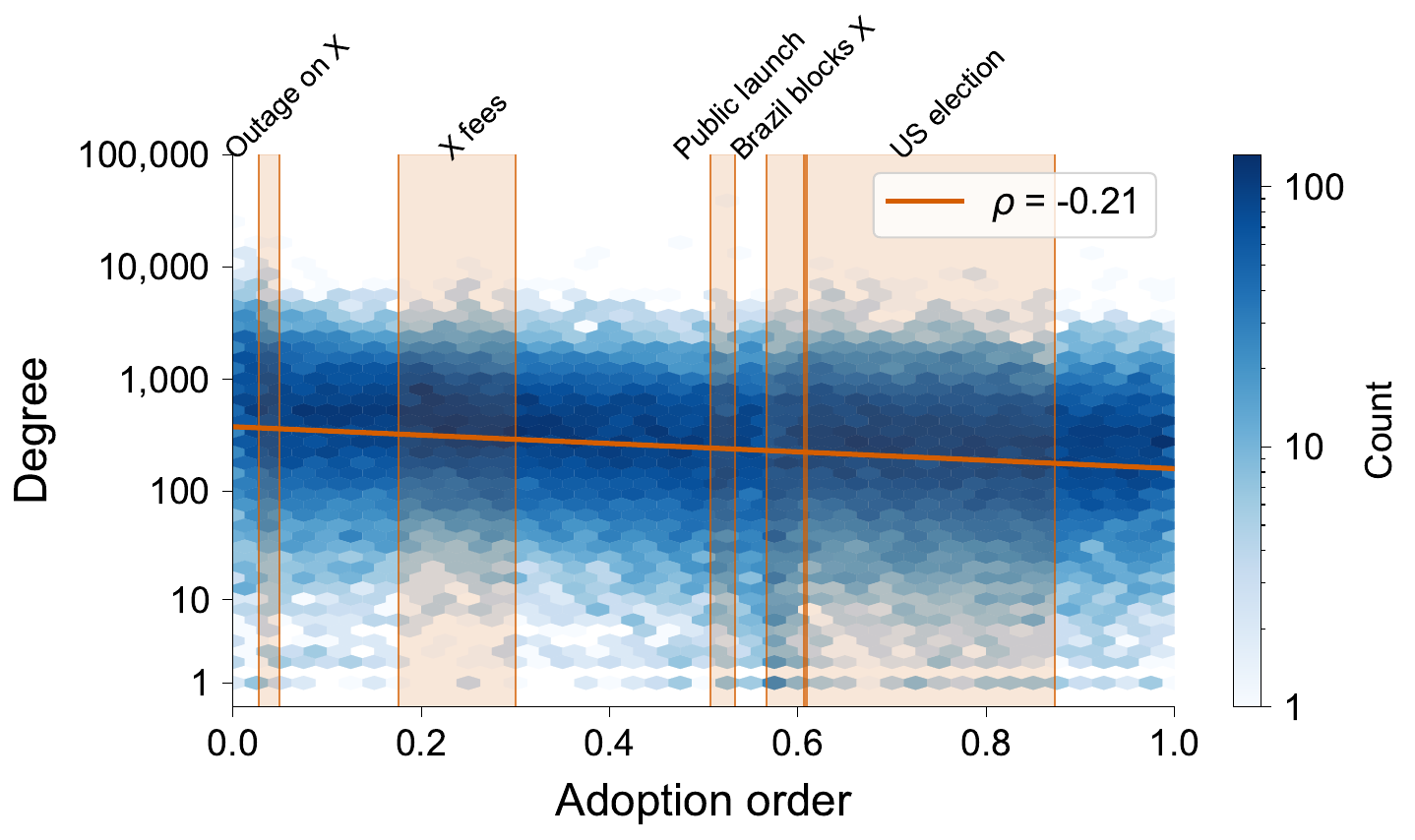}
  \caption{\textbf{Degree-adoption order correlation test from\citeauthor{cencettiDistinguishingSimpleComplex2023}\autocite{cencettiDistinguishingSimpleComplex2023}}. The observed negative correlation ($\rho = -0.21$, $p < 0.001$, $N = 49{,}533$) is consistent with simple contagion as the dominant mechanism.}
  \label{fig:S_cencetti}
\end{figure}

Additionally, we apply the test proposed by \citeauthor{cencettiDistinguishingSimpleComplex2023}\autocite{cencettiDistinguishingSimpleComplex2023} to distinguish simple from complex contagion based on network topology and adoption timing. Under simple contagion, high-degree nodes are reached early, and a cascade proceeds to low-degree nodes, producing a negative correlation between degree and adoption order. Under complex contagion requiring threshold reinforcement, this relationship weakens or reverses. Figure \ref{fig:S_cencetti} shows that the correlation between degree and adoption order is negative ($\rho = -0.21$, $p < 0.001$, $N = 49{,}533$), which is consistent with simple contagion as the dominant mechanism. This test provides independent validation of our ML classifier results using only structural network information rather than simulated cascade features.

\printbibliography
\end{document}